\documentclass[%
 reprint,
 superscriptaddress,
 amsmath,amssymb,
 prx,
 floatfix,
]{revtex4-2}

\usepackage{graphicx}
\usepackage{dcolumn}
\usepackage{bm}
\usepackage{microtype}
\usepackage{dsfont}
\usepackage{mathtools}
\usepackage{comment}
\usepackage{hyperref}

\usepackage[normalem]{ulem}

\hypersetup{colorlinks=true}

\usepackage{color} 

\def\Xint#1{\mathchoice
   {\XXint\displaystyle\textstyle{#1}}%
   {\XXint\textstyle\scriptstyle{#1}}%
   {\XXint\scriptstyle\scriptscriptstyle{#1}}%
   {\XXint\scriptscriptstyle\scriptscriptstyle{#1}}%
   \!\int}
\def\XXint#1#2#3{{\setbox0=\hbox{$#1{#2#3}{\int}$}
     \vcenter{\hbox{$#2#3$}}\kern-.5\wd0}}
\def\ddashint{\Xint=}

\DeclareMathOperator*{\SumInt}{%
\mathchoice%
  {\ooalign{$\displaystyle\sum$\cr\hidewidth$\displaystyle\int$\hidewidth\cr}}
  {\ooalign{\raisebox{.14\height}{\scalebox{.7}{$\textstyle\sum$}}\cr\hidewidth$\textstyle\int$\hidewidth\cr}}
  {\ooalign{\raisebox{.2\height}{\scalebox{.6}{$\scriptstyle\sum$}}\cr$\scriptstyle\int$\cr}}
  {\ooalign{\raisebox{.2\height}{\scalebox{.6}{$\scriptstyle\sum$}}\cr$\scriptstyle\int$\cr}}
}

\definecolor{myred}{rgb}{0, 0, 0}

\begin{document}

\preprint{APS/123-QED}

\title{Exact Continuum Representation of Long-range Interacting Systems \\ \textcolor{myred}{and Emerging Exotic Phases in Unconventional Superconductors}}

\author{Andreas A. Buchheit}
 \email{buchheit@num.uni-sb.de}
  \affiliation{%
 Saarland University, 66123 Saarbr{\"u}cken, Germany
}%
\author{Torsten Ke{\ss}ler}%
\affiliation{%
 Saarland University, 66123 Saarbr{\"u}cken, Germany
}%
\affiliation{
Eindhoven University of Technology, 5600 MB Eindhoven, Netherlands
}
 \author{Peter K. Schuhmacher}
\author{Benedikt Fauseweh}%
 \affiliation{%
 German Aerospace Center (DLR), 51147 Cologne, Germany
}%

\date{\today}
             
\begin{abstract}
Continuum limits are a powerful tool in the study of many-body systems, yet their validity is often unclear when long-range interactions are present. In this work, we rigorously address this issue and put forth an exact representation of long-range interacting lattices that separates the model into a term describing its continuous analog, the integral contribution, and a term that fully resolves the microstructure, the lattice contribution. 
For any system dimension, any lattice, any power-law interaction, and for linear, nonlinear, and multi-atomic lattices, we show that the lattice contribution can be described by a differential operator based on the multidimensional generalization of the Riemann zeta function, namely the Epstein zeta function. \textcolor{myred}{
We employ our representation in Fourier space to solve the important problem of long-range interacting unconventional superconductors. We derive a generalized Bardeen--Cooper--Schrieffer gap equation and find emerging exotic phases in two-dimensional superconductors with topological phase transitions. Finally, we utilize non-equilibrium Higgs spectroscopy to analyze the impact of long-range interactions on the collective excitations of the condensate. We show that the interactions can be used to fine-tune the Higgs mode's stability, ranging from exponential decay of the oscillation amplitude up to complete stabilization. }
\end{abstract}

\maketitle

\section{Introduction}

Long-range interactions are ubiquitous in nature across all scales. Such interactions are of fundamental importance in all of physics, e.g., long-range Coulomb interactions leading to the formation of Bose--Einstein condensates in cold atoms \cite{PhysRevLett.84.5687}, the speed of correlation propagation in long-range interacting Ising systems in trapped ion quantum simulations \cite{Richerme2014}, dipolar interactions between spins in spin-ice materials \cite{Castelnovo2008}, and other phenomena in nanoscale systems \cite{RevModPhys.82.1887}.
They are the driver behind the formation of complex structures, from the quarks that form the atomic nucleus over the microscopic formation of solids and molecules based on atoms and ions to galaxy patterns spanning billions of light years.

Modeling and predicting the emergent dynamics of systems that are subject to such long-range interactions requires the computation of the interaction energy. For many-body systems on lattices, this task becomes a problem for numerical approaches, as the computational effort scales directly with the number of particles involved, making calculations for macroscopic systems, e.g., $N=10^{23}$ atoms, impossible.
In special cases, tricks like Ewald summation~\cite{ewald1921berechnung}
or the reorganization of the sum~\cite{madelung1918elektrische, marathe1983electrostatic, wolf1999exact} yield converging alternative formulations of the original lattice sum.
However, in the general case, these methods are not applicable
and give limited information about the important analytic properties of the sum.
One approach to solve this problem is the continuum limit, in which the lattice spacing is taken to zero, and sums can be replaced by computable integrals. In the context of quantum mechanical systems, this procedure corresponds to identifying the effective field theory that describes the low energy excitation spectrum of the lattice system \cite{Essler_Massive_Quantum_Field}. Also, the inverse task, i.e., deriving a lattice theory from a field theory, is of practical relevance for strongly interacting lattice gauge theories in high-energy physics \cite{Ratti_2018}. 
While the continuum limit is a powerful tool in theoretical physics, it must be yielded with caution in systems with long-range interactions, e.g., that decay as a power law with distance. 
When applying continuum renormalization schemes in such systems, it is typically observed that the long-range interaction leads to divergences in the resulting flow equations \cite{maghrebi2017continuous}. This requires a change in the applied methods or even presents a fundamental hurdle that prohibits the use of the continuum limit \cite{Dubin1997}. In many cases, artificial cutoff energies need to be introduced that have to be justified in hindsight. 

The goal of this work is to address this issue by fundamentally changing our understanding of how discrete and continuous systems with long-range interactions are related. In contrast to standard continuum approximations, we put forth a continuum representation of the discrete lattice that is exact, systematic, and parameter-free. We show that the discrete lattice problem can be separated into a term that describes its continuous analog, the continuum contribution, and a term that includes all information about the microstructure, the lattice contribution, hence demonstrating equivalence between lattice and continuum.
 To this end, we apply the recently developed Singular Euler--Maclaurin (SEM) expansion~\cite{buchheit2022efficient,buchheit2022singular,thesis_buchheit_2021}, which generalizes the 300-year old Euler--Maclaurin summation formula, and extend it to nonlinear and multi-atomic systems. The singular lattice sum is expressed in terms of an integral and a lattice contribution described by a differential operator, both of which are efficiently computable. 
Performing a scaling analysis, we determine the circumstances under which the lattice contribution is of particular relevance. Among others, we show that the correction becomes quasi-scale-invariant and hence remains important at all scales if the interaction exponent is equal to the system dimension.

\textcolor{myred}{Our continuum representation yields an efficient numerical method and an analytic toolset for simulating and understanding long-range interacting systems}. We demonstrate the performance of this method by investigating three highly important yet highly challenging physical examples: In Example 1, we study dipolar interactions of Skyrmions in a two-dimensional spin lattice. Example 2 shows that our method readily extends to nonlinear systems. Here we study the full nonlinear Coulomb forces in an ion chain with topological defects. Our method yields a nonlinear sine-Gordon model with long-range interactions, where we obtain analytical results for the lattice contribution. In Example 3, we analyze spin waves in a three-dimensional pyrochlore lattice with dipole interactions, which are notoriously difficult to compute~\cite{bramwell2001spin}. In all cases, our method proves to be both highly accurate and fast, whereas the continuum approximation fails either on a quantitative (Examples 1 and 2) or even on a qualitative level (Example 3).

\textcolor{myred}{In the second part of this work, we apply our representation in Fourier space to quantum lattices and solve the important problem of understanding long-range interactions in unconventional superconductors. Starting from a single band model with long-range density-density interactions, we derive a generalization of the Bardeen--Cooper--Schrieffer (BCS) gap equation for triplet superconductors that is valid for general power-law electron-electron interactions and exactly describes the full microstructure of the material within the mean field approximation. We solve the generalized gap equation for two-dimensional superconductors, determine the ground state, and show that additional exotic phases emerge in the phase diagram due to the long-range interaction, one being topologically nontrivial. We then evaluate the non-equilibrium dynamics of the condensate after a sudden quench and the impact of the long-range interactions on the properties of the arising Higgs oscillations. Here, the decay of the Higgs mode's amplitude can be either accelerated or completely avoided depending on the range of the interaction. We provide the full implementation of the three benchmark examples as well as the data analysis for the long-range superconductor in Mathematica in the supplementary material and on GitHub~\cite{Buchheit_Continuum_representation_2022}.}

This work is structured as follows: In Sec.~\ref{sec:linear_systems}, we derive the new representation and apply it to the study of Skyrmions.
Sec.~\ref{sec:nonlinear_systems} subsequently generalizes our representation to nonlinear systems illustrated by the analysis of defects in an ion chain. 
In Sec.~\ref{sec:multi_atomic}, we extend the method to multi-atomic lattices with an application to spin waves in a three-dimensional pyrochlore lattice. 
\textcolor{myred}{Finally, we solve the problem of long-range interacting unconventional superconductors in Sec.~\ref{sec:unconventional_superconductivity}.} 
We draw our conclusions and offer an outlook into further applications and extensions of our work in Sec.~\ref{sec:conclusions}. 

\section{Representation for linear systems}
\label{sec:linear_systems}
We consider a lattice $\Lambda = A_\Lambda \mathds Z^d$ in $d$ dimensions of identical discrete constituents in the most general sense, be it atoms, molecules, spins, or states, where the columns of the regular matrix $A_\Lambda\in\mathds R^{d\times d}$ are the lattice vectors. We will refer to these discrete constituents as atoms in the following. These atoms shall interact via a long-range power-law potential 
\[
s(\bm y)=\vert \bm y \vert^{-\nu}
\] 
with arbitrary exponent $\nu\in \mathds C$.
Our goal is now to find a continuum representation of this discrete long-range interacting system that captures the effect of its inherent discreteness.

\subsection{The Singular Euler--Maclaurin expansion}
\label{subsec:sem_derivation}
We start by computing the interaction energy $U$ of a test particle at position $\bm x\in \mathds R^d$ with all atoms of the lattice inside a  (typically unbounded) region $\Omega$. It reads
\[
    U(\bm x)= \sideset{}{'}\sum_{\bm y \in \Omega \cap \Lambda} \frac{g_{\bm y}}{\vert \bm y -\bm x \vert^\nu}.
\]
Here the primed sum excludes the self-energy term $\bm y=\bm x$ in case the test particle belongs to the lattice.
The  placeholder $g_{\bm y}$ describes the state of the lattice atom at position $\bm y$. For example, $g_{\bm y}$ could be a displacement from an equilibrium position in an atomic crystal, or in the case of spins, a scalar product of spin orientations $g_{\bm y}=\bm S_{\bm x}\cdot \bm S_{\bm y}$. If the quantity $g_{\bm y}$ varies sufficiently slowly in $\bm y$, then it is natural to replace the discrete values by their interpolation 
$g_{\bm y}\to g(\bm y)$, with $g$ smooth and sufficiently band-limited (its Fourier transform is concentrated in the first Brillouin zone), and subsequently try to approximate the discrete lattice sum by a related integral. Care has to be taken, as the singularity is not necessarily integrable; hence a regularization is required. One possibility is to remove an $\varepsilon$-ball around $\bm x$ from integration, which corresponds to a standard ultraviolet cutoff~\cite{wilson1974confinement}. We then have
\begin{align}
    U(\bm x)= \mathcal I_\varepsilon(\bm x)+\mathcal Z_\varepsilon(\bm x)
    \label{eq:SEM_epsilon}
\end{align}
with 
\[
    \mathcal I_\varepsilon(\bm x)=\frac{1}{V_\Lambda} \int  \limits_{\Omega\setminus B_\varepsilon(\bm x)} \frac{g(\bm y)}{\vert \bm y-\bm x\vert^\nu}\, \mathrm d \bm y,
\]
with $V_\Lambda=\vert\det A_\Lambda\vert$  the volume of the elementary lattice cell and $\mathcal Z_\varepsilon$ the lattice contribution \footnote{In this manuscript, calligraphic symbols carry a dependency on $\Lambda$ and $\nu$ where we avoid explicit indexing.}. 
Another option is to use the Hadamard regularization,
see Appendix~\ref{sec:appendix_hadamard}, in which case 
\begin{align}
    U(\bm x)= \mathcal I(\bm x)+\mathcal Z(\bm x),
    \label{eq:sem_hadamard}
\end{align}
where
\[
\mathcal I(\bm x)=\frac{1}{V_\Lambda} \ddashint \limits_{\Omega} \frac{g(\bm y)}{\vert \bm y-\bm x\vert^\nu}\, \mathrm d \bm y
\]
with the lattice contribution $\mathcal Z$.
Neglecting the lattice contribution leads to the standard integral approximation often used in condensed matter physics, see e.g.~Ref~\cite{maghrebi2017continuous}. \textcolor{myred}{The Hadamard integral, denoted by the dashed integral sign, is the natural extension of the standard Lebesgue integral to functions with power-law singularities \cite{gelfand1964generalizedI}. If the function remains integrable, then the Hadamard integral coincides with the integral's usual definition. For non-integrable power-law singularities with $\mathrm{Re}(\nu)\le d$, the Hadamard integral forms the analytic continuation in the exponent $\nu$.}  

The goal of this work is to quantify the lattice contribution $\mathcal Z$ (resp. $\mathcal Z_\varepsilon$) in all generality for any number of spatial dimensions and any lattice. \textcolor{myred}{We will show that it is possible to cast the lattice contributions in terms of derivatives of $g$,
\begin{equation*}
 \mathcal Z(\bm x)=\mathcal D g(\bm x)
\end{equation*}
with $\mathcal D$ a differential operator of infinite order that can be suitably truncated and whose coefficients can be efficiently computed.}

Our analysis will reveal that this correction is highly relevant in many physical systems of interest, even if a slowly varying $g$ suggests that a continuum approximation is appropriate. \textcolor{myred}{The lattice contribution becomes particularly important  if the interaction exponent $\nu$ approaches the system dimension $d$. In this work, we demonstrate that in this regime, our continuum representation offers qualitative insights that the standard integral approximation cannot provide, as it either becomes ill-defined or exhibits uncontrollable errors.}

For clarity of presentation, we focus on $\mathcal Z$ and show how to obtain $\mathcal Z_\varepsilon$ later on. We set out by writing $\mathcal Z$ as the difference between a discrete and a continuous system,
\[
\mathcal Z(\bm x)= \sideset{}{'}\sum_{\bm y \in \Omega \cap \Lambda} f_{\bm x}(\bm y) - \frac{1}{V_\Lambda} \ddashint \limits_{\Omega} f_{\bm x}(\bm y)\, \mathrm d \bm y=\SumInt_{\bm y \in \Omega,\Lambda} f_{\bm x}(\bm y),
\]
with $f_{\bm x}(\bm y)=g(\bm y)/\vert \bm y-\bm x\vert^\nu$.
As such differences will reappear often in our considerations, it is useful to introduce the corresponding operator on the right-hand side, which is called the sum-integral~\cite{buchheit2022singular}.
In this work, we focus on the case of an infinite lattice $\Omega=\mathds R^d$ to avoid additional geometry-dependent contributions due to boundaries.
The inclusion of boundary effects is planned for a forthcoming publication.

We now show that the lattice contribution can be written in terms of a differential operator, which acts on the smooth function $g$ only, and whose coefficients include the interaction potential and the lattice structure. 
The following steps are based on the key idea of restricting the range of the interaction potential $s$ by introducing an exponentially decaying cutoff function
$e^{-\beta \vert \bm y\vert^2}$, $\beta>0$. \textcolor{myred}{Physically, the range of the long-range interaction is restricted to a length scale $1/\sqrt{\beta}$, rendering it effectively short-ranged.
Subsequently, the original true long-range interaction is restored by taking the limit $\beta \to 0$ outside of  sum and integral,  }
\[
\mathcal Z(\bm x)=\lim_{\beta\to 0}\SumInt_{\bm y \in \mathds R^d,\Lambda} e^{-\beta \vert \bm y\vert^2}\frac{g(
 \bm y)}{\vert \bm  y-\bm x\vert^\nu}.
\]
This procedure avoids divergent terms later on and guarantees convergence of the arising Dirichlet series. Indeed, for $g=P$ a polynomial of arbitrary degree, a key result of Ref.~\cite{buchheit2022singular} shows that 
\begin{equation}\label{eq:dirichlet-series-beta-regularization}
\lim_{\beta\to 0}\SumInt_{\bm y \in \mathds R^d,\Lambda} e^{-\beta \vert \bm y\vert^2}\frac{P(
 \bm y)}{\vert \bm  y-\bm x\vert^\nu} = \sideset{}{'} \sum_{\bm y\in \Lambda}\frac{P(
 \bm y)}{\vert \bm  y-\bm x\vert^\nu},
\end{equation}
if the Dirichlet series converges a priori without $\beta$-regularization. If the regularization is required, then the  sum-integral creates the meromorphic continuation of the right-hand side in $\nu$. 
  
With this result and for $g$ sufficiently differentiable, we can now expand $g$ in a Taylor series around $\bm x$ of order $2\ell+1$. The cutoff function allows us to exchange the sum due to the Taylor series with the sum-integral and the $\beta$-limit \footnote{This is a nontrivial result, for details see \cite{buchheit2022singular}.}, resulting in a representation of the lattice contribution in terms of derivatives of $g$ 
\begin{equation}
 \mathcal Z(\bm x)=\mathcal D g(\bm x)=\mathcal D^{(\ell)} g(\bm x)+\mathcal O(\Delta^{\ell+1} g),
\label{eq:SEM_expansion}
\end{equation}
with $\mathcal D$ a differential operator of infinite order, $\mathcal D^{(\ell)}$ its truncation up to order $2\ell+1$, and $\Delta$ the Laplacian. 
The representation of the lattice contribution in Eq.~\eqref{eq:SEM_expansion} is called the Singular Euler-Maclaurin (SEM) expansion, a full derivation of which is provided in \cite{buchheit2022efficient,buchheit2022singular,thesis_buchheit_2021}.
For $\Omega = \mathbb R^d$ the SEM operator $\mathcal D^{(\ell)}$ takes the particularly simple form
\begin{equation}
    \mathcal D^{(\ell)}=\sum_{k=0}^{2\ell+1} \frac{1}{k!} \sideset{}{'} \sum_{\bm y\in (\Lambda-\bm x)} \frac{(\bm y
   \cdot \bm \nabla)^{k}}{\vert \bm y\vert^\nu},
   \label{eq:sem_operator}
\end{equation}
where the lattice sums are to be understood in the sense of Eq.~\eqref{eq:dirichlet-series-beta-regularization},
i.e., the lattice sum is replaced by the value of the meromorphic continuation if it does not converge in the classical sense.
In the following, we show how the coefficients of this operator can be efficiently evaluated for lattices in arbitrary dimensions.

\subsection{Representation in terms of Epstein zeta}
\label{subsec:epstein_zeta}
We demonstrate that the operator coefficients can be obtained from an efficiently computable generalization of the Riemann zeta function to higher dimensions, the Epstein zeta function $Z_{\Lambda,\nu}$ for the lattice $\Lambda$ and the exponent $\nu$. It reads \cite{epstein1903theorieI,epstein1903theorieII,borwein2013lattice}
  \begin{equation*}
 Z_{\Lambda,\nu}\left\vert \begin{matrix}
      \bm x\\\bm y
    \end{matrix}\right\vert=\,\sideset{}{'}\sum_{\bm z \in \Lambda} \frac{e^{-2\pi i  \bm y \cdot \bm z}}{ {\vert \bm z-\bm x\vert}^{\nu}}.
  \end{equation*}
  The Epstein zeta function has been used, among others, by Emersleben in the study of ionic crystal potentials in Refs.~\cite{emersleben1923zetafunktionenI,emersleben1923zetafunktionenII}.
  The function is smooth in $\bm y$ outside points of the reciprocal lattice $\Lambda^\ast=(A_\Lambda^{-1})^T\mathds Z^d$ where it exhibits singularities that are described by the Fourier transform of the interaction $s(\bm y)=\vert \bm y \vert^{-\nu}$. We subsequently subtract the singularity at $\bm y=\bm 0$ and define the regularized function 
   \begin{equation}
    Z_{\Lambda,\nu}^{\text{reg}}\left\vert \begin{matrix}
      \bm x\\\bm y
    \end{matrix}\right\vert=e^{2\pi i \bm x\cdot \bm y}Z_{\Lambda,\nu}\left\vert \begin{matrix}
      \bm x\\\bm y
    \end{matrix}\right\vert -\frac{\hat s(\bm y)}{V_\Lambda}.
    \label{eq:z_lambda_reg}
  \end{equation}
  Here the Fourier transform of the interaction reads
  \begin{equation}\label{eq:fourier-interaction}
   \hat s(\bm y)= \pi^{\nu-d/2}\frac{ \Gamma ((d-\nu)/2)}{\Gamma(\nu/2) } \vert\bm y\vert^{\nu-d},   
  \end{equation} see e.g.~\cite[p.~349]{gelfand1964generalizedI}. The function $ Z^{\text{reg}}_{\Lambda,\nu}$ is analytic in $\bm y$ around zero and allows us to compute analytic continuations of lattice sums by means of derivatives in $\bm y$, namely
  \[
    \,\sideset{}{'}\sum_{\bm z \in (\Lambda-\bm x)} \frac{P(\bm z)}{ {\vert \bm z\vert}^{\nu}}=P\bigg(\frac{i \bm \nabla_{\bm y}}{2\pi }\bigg)Z_{\Lambda,\nu}^{\text{reg}}\left\vert \begin{matrix}
      \bm x\\\bm y
    \end{matrix}\right\vert \Bigg\vert_{\bm y=0}.
  \]
  In particular, $ Z_{\Lambda,\nu}^{\text{reg}}$ reduces to $Z_{\Lambda,\nu}$ for $\bm y= \bm 0$.   These lattice sums define the coefficients of the SEM operator in Eq.~\eqref{eq:sem_operator}.
  The infinite order SEM operator hence can be cast as
  \begin{equation}
    \mathcal D g(\bm x)=Z_{\Lambda,\nu}^{\text{reg}}\left\vert \begin{matrix}
      \bm x\\\frac{i\bm \nabla}{2\pi }
    \end{matrix}\right \vert \,g(\bm x),
  \end{equation}
  in the sense of a Taylor expansion of $Z_{\Lambda,\nu}^{\text{reg}}$ in its second argument around zero, and where the gradient only acts on $g$. In this way, derivatives of the interaction potential $s$ are avoided, which rapidly increase in size with the derivative order, and which would result in the divergence of the standard Euler-Maclaurin summation formula \cite{apostol1999euler}. The infinite order SEM expansion of the potential energy $U$ then yields the continuum representation of the discrete lattice
  \begin{equation}
    U(\bm x) = \frac{1}{V_\Lambda} \ddashint \limits_{\mathds R^d} \frac{g(\bm y)}{\vert \bm y-\bm x\vert^\nu}\, \mathrm d \bm y+Z_{\Lambda,\nu}^{\text{reg}}\left\vert \begin{matrix}
      \bm x\\\frac{i\bm \nabla}{2\pi }
    \end{matrix}\right \vert  \,g(\bm x),
    \label{eq:sem_expansion}
  \end{equation}
  and the SEM expansion of order $\ell$ is obtained by replacing $\mathcal D$ by $\mathcal D^{(\ell)}$ with an error that scales as $\Delta^{\ell+1}g$. Note that Eq.~\eqref{eq:sem_expansion} is exact and involves no approximation. Here, the integral models the interaction of the test particle with a continuum. Hence, the inherent discreteness of the lattice is completely captured by the second term. Among others, the distance of the test particle to the nearest lattice atom is included in the lattice contribution. This contribution becomes, among others, particularly relevant if the test particle approaches a lattice atom.
  
As there exist exponentially convergent series representations for $Z_{\Lambda,\nu}$~\cite{crandall2012unified}, and hence for $Z_{\Lambda,\nu}^{\text{reg}}$, for any number of space dimensions, the lattice contribution can be efficiently computed. We provide an efficient implementation of $Z_{\Lambda,\nu}^{\text{reg}}$ for lattices in an arbitrary number of space dimensions along with this article ~\cite{Buchheit_Continuum_representation_2022}. 

\subsection{Alternative regularizations of the interaction}
\label{subsec:interaction_regularizations}
So far, we have investigated the lattice contribution $\mathcal Z$ where the integral has been made well-defined by means of the Hadamard regularization. We now investigate alternative regularizations where a short-range cutoff of the interaction is applied. Here, the regularized interaction $s_\varepsilon$ coincides with $s$ outside of an $\varepsilon$-ball,
\[
    s_\varepsilon(\bm y) = s(\bm y),\quad \vert \bm y\vert\ge \varepsilon,
\]
and the interaction is replaced by an arbitrary integrable function for $\vert\bm y\vert< \varepsilon$. Then the corresponding lattice contribution $\mathcal Z_\varepsilon$ reads
\begin{equation}
\mathcal Z_\varepsilon(\bm x)= Z_{\Lambda,\nu,\varepsilon}^{\text{reg}}\left\vert \begin{matrix}
      \bm x\\\frac{i\bm \nabla}{2\pi}
    \end{matrix}\right\vert
    \label{eq:SEM_operator_explicit}
\end{equation}
where the function $Z_{\Lambda,\nu,\varepsilon}^{\text{reg}}$ is obtained by replacing $\hat s$ by $\hat s_\varepsilon$ in Eq.~\eqref{eq:z_lambda_reg}. For the special case of a hard cutoff, where $s_\varepsilon = 0$ inside the $\varepsilon$-ball, we have 
\begin{equation*}
Z_{\Lambda,\nu,\varepsilon}^{\text{reg}}\left\vert \begin{matrix}
      \bm x\\\bm y
    \end{matrix}\right\vert =Z_{\Lambda,\nu}^{\text{reg}}\left\vert \begin{matrix}
      \bm x\\\bm y
    \end{matrix}\right\vert+\frac{1}{V_\Lambda} \ddashint \limits_{B_\varepsilon} \frac{e^{-2\pi i \bm y\cdot \bm z}}{\vert \bm z\vert^\nu}\,\mathrm d\bm z.
\end{equation*}
The Hadamard integral on the right can be expanded in the following way
\begin{equation}
    \frac{\omega_d}{V_\Lambda}\sum_{k=0}^{\infty}  \frac{ (1/2)_{k}}{(2k)!(d/2)_{k}} \frac{\varepsilon^{2k+d-\nu}}{2k+d-\nu} (2\pi i \bm y)^{2k},
    \label{eq:integral_expansion}
\end{equation}
with $\omega_d$ the surface area of the sphere in $d$ dimensions and where $(x)_k$ is the Pochhammer symbol.
Note that for any $s_\varepsilon$, the new function $Z_{\Lambda,\nu,\varepsilon}^{\text{reg}}$ is entire in $\nu$. Thus formula \eqref{eq:SEM_operator_explicit} holds for all interaction exponents. 
  
\subsection{Quasi scale-invariant lattice contributions}
\label{subsec:scaling}
After having shown how to describe the lattice contribution in the most general way by means of the SEM expansion, we discuss under which circumstances it is relevant. To this end, we first fix the position $\bm x$ of the test particle in space. We then perform a scale transformation of $\bm g$ around $\bm x$, setting 
\[
    g_\lambda(\bm y) = g(\bm x+ (\bm y-\bm x)/\lambda)
\] 
with a scaling factor $\lambda>1$. The rescaled function $g_\lambda$ now varies at a characteristic length scale proportional to $\lambda$, its bandwidth scales as $\lambda^{-1}$, and it coincides with $g$ at the position of the test particle $\bm x$. We now define the potential energy under scale transformation $U_\lambda$ by the replacement $g\to g_\lambda$,
\[
U_\lambda(\bm x)=\sideset{}{'}\sum_{\bm y\in \Lambda} \frac{g_\lambda(\bm y)}{\vert \bm y-\bm x \vert^\nu}.
\]
Subsequently, we can choose between two options for using the SEM in order to divide $U_\lambda$  into a term that describes the continuum approximation of the system and a part that describes the lattice contribution, namely  Eqs.~\eqref{eq:SEM_epsilon}-\eqref{eq:sem_hadamard}. We can either exclude an $\varepsilon$-ball from integration, or we can make use of the Hadamard regularization. We first discuss the Hadamard regularization where
\begin{align*}
    U_\lambda(\bm x)= \mathcal I[g_\lambda](\bm x)+\mathcal Z[g_\lambda](\bm x).
\end{align*}
The scaling of the Hadamard integral with $\lambda$ then follows as
\[
\mathcal I[g_\lambda](\bm x)=\lambda^{d-\nu} \mathcal I(\bm x),
\]
and a Taylor expansion in the lattice contribution yields 
\[
\mathcal Z[g_\lambda](\bm x)=Z_{\Lambda,\nu}^{}\left\vert \begin{matrix}
      \bm x\\\bm 0
    \end{matrix}\right\vert g(\bm x)+\mathcal O(\lambda^{-1}).
\]
The potential energy thus obeys the scaling law 
\[
U_\lambda(\bm x)=\lambda^{d-\nu} \mathcal I(\bm x) +Z_{\Lambda,\nu}^{}\left\vert \begin{matrix}
      \bm x\\\bm 0
    \end{matrix}\right\vert g(\bm x)+\mathcal O(\lambda^{-1}).
\]
We conclude that for strong long-range interactions with $\mathrm{Re}(\nu)<d$, the integral scales as $\lambda^{d-\nu}$ and hence dominates the lattice contribution that converges to a constant for $\lambda\to \infty$. In this case, the lattice contribution remains relevant for  systems with mesoscopic $\lambda$, or if high precision is required. On the other hand, for $\text{Re}(\nu)>d$, the lattice contribution is the dominating quantity, and the interaction is thus effectively short-ranged. \textcolor{myred}{Note that the scaling observed here cannot be removed by standard techniques for systems with super-extensive energies such as Kac rescaling. The quantities considered here are intensive and hence well-defined in the thermodynamic limit of infinite particles.}

We now investigate the scaling in case the $\varepsilon$-cutoff is used,
\begin{align*}
    U_\lambda(\bm x)= \mathcal I_\varepsilon[g_\lambda](\bm x)+\mathcal Z_\varepsilon[g_\lambda](\bm x).
\end{align*}
The integral can then be rewritten as 
\[
   \mathcal I_\varepsilon[g_\lambda](\bm x)=\lambda^{d-\nu} \mathcal I_{\varepsilon/\lambda}(\bm x).
\]
We subsequently divide the integration region into the cases $\vert \bm y-\bm x\vert>\varepsilon$ and $\varepsilon/\lambda<\vert\bm y-\bm x\vert<\varepsilon$,
\[
    \mathcal I_{\varepsilon/\lambda}(\bm x)=\mathcal I_{\varepsilon}(\bm x)+\frac{1}{V_\Lambda}\int \limits_{\varepsilon/\lambda<\vert \bm y\vert<\varepsilon} \frac{g(\bm y+\bm x)}{\vert \bm y\vert^\nu}\,\mathrm d\bm y.
\]
We then find after expanding $g$ on the right hand side around $\bm x$ and using Eq.~\eqref{eq:integral_expansion} that
\begin{align*}
    &\mathcal I_\varepsilon[g_\lambda](\bm x)=\lambda^{d-\nu}\mathcal I_\varepsilon(\bm x)\notag \\
&+\frac{\omega_d}{V_\Lambda}\sum_{k=0}^\infty \frac{(1/2)_k}{(2k)!(d/2)_k}\frac{\varepsilon^{d+2k-\nu}}{\lambda^{2k}}\frac{\lambda^{d+2k-\nu}-1}{(d+2k)-\nu}\Delta^k g(\bm x),
\end{align*}
where, in case that $\nu=2k+d$, we note that
\[
    \lim_{\nu\to 2k+d}\frac{\lambda^{d+2k-\nu}-1}{(d+2k)-\nu}=\log\lambda.
\]
Hence, we obtain the scaling 
\[
    \mathcal I_\varepsilon[g_\lambda](\bm x)=\left\{\begin{matrix}\mathcal O(\lambda^{d-\nu})+\mathcal O(\lambda^0),\quad \nu \neq d, \\ \mathcal O(\lambda^0)+\mathcal O(\log \lambda),\quad \nu = d.
\end{matrix} \right.
\]
The lattice contribution for finite $\varepsilon$ yields
\begin{align*}
&\mathcal Z_\varepsilon[g_\lambda](\bm x)=Z_{\Lambda,\nu,\varepsilon}^{\text{reg}}\left\vert \begin{matrix}
      \bm x\\\frac{i \bm \nabla}{2\pi \lambda}
    \end{matrix}\right\vert g(\bm x).
\end{align*}

As in the case of the Hadamard regularization, the integral dominates for $\text{Re}(\nu)<d$ (strong long-range interactions). In contrast to that, for $\text{Re}(\nu)>d$, both continuum and lattice contribution include scale-invariant terms. Hence both remain relevant. In the limiting case $\nu=d$, the continuum contribution scales as $\log \lambda$ and is hence of the same order of magnitude as the lattice contribution, even in the case of macroscopic $\lambda$. In this highly relevant scenario, the lattice contribution needs to be taken into account at all scales, even in the thermodynamic limit, in order to obtain results that are qualitatively reliable. We call these lattice contributions quasi-scale-invariant as the error of the continuum approximation only decreases logarithmically with the scale $\lambda$ and cannot be assumed small even at macroscopic scales.

\subsection{Example 1: Skyrmions in a spin lattice}
\label{subsec:skyrmions}

In order to illustrate the performance of our method, we now study dipolar interactions in a $d=2$ spin lattice with two interacting Skyrmions. In recent years, the study of Skyrmions, topologically protected quasi-particles in spin lattices, has gained significant attention, see the reviews in Refs.~\cite{wiesendanger2016nanoscale,bogdanov2020physical,tokura2020magnetic}. As stable Skyrmions at room temperature have been observed \cite{das2019observation,jiang2017direct,gilbert2015realization}, and as methods for creating, deleting, and manipulating them have been developed \cite{romming2013writing,litzius2020role,hrabec2017current}, they are considered as promising candidates for storing and manipulating information in novel spintronics devices \cite{wiesendanger2016nanoscale,kang2016skyrmion,li2021magnetic}. Recently, a Skyrmion Hall effect has been observed \cite{chen2017skyrmion,jiang2017direct}, offering a new way for manipulating these quasi-particles. Skyrmions have been proposed as a platform for neuromorphic computing \cite{song2020skyrmion} and as qubits for quantum computing \cite{psaroudaki2021skyrmion}. Quantum effects in Skyrmion systems have been investigated \cite{lohani2019quantum,janson2014quantum}. Recently, it has been conjectured that the long-range dipole interaction is relevant for a correct quantitative description of their behavior \cite{zhang2017skyrmion,jena2020elliptical,schwarze2015universal}.

\begin{figure}
    \centering
    \includegraphics[width=.52\textwidth]{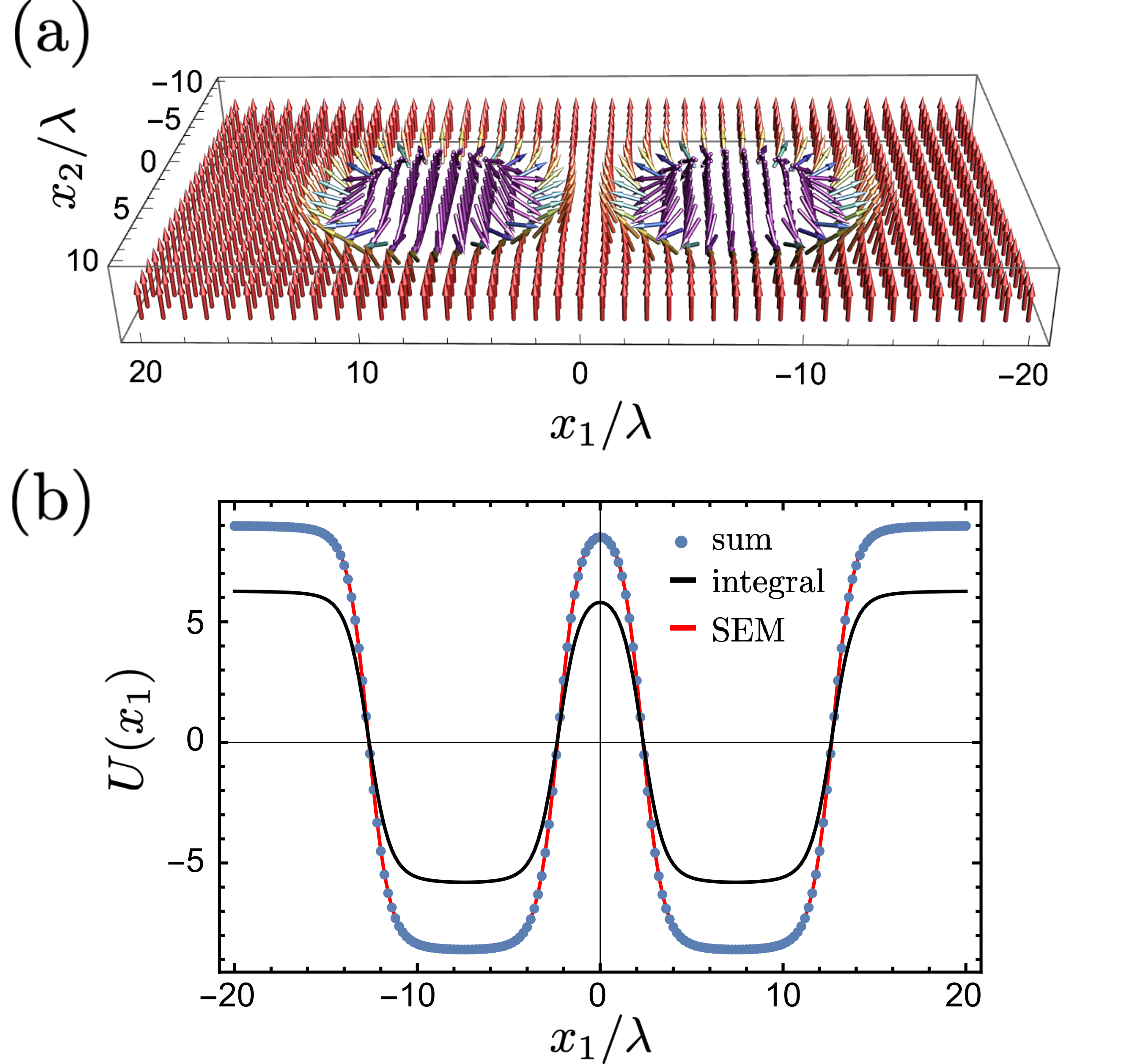}
    \caption{(a) 2D spin lattice with two Néel Skyrmions. The Skyrmions have identical domain wall width $\lambda=5$,  radius $26/5\,\lambda$, and are separated by $15\,\lambda$.  (b) Potential energy $U$ for a central spin $\bm S_c=\bm e_3$ as a function of $x_1$ for $x_2=0$. The exact energies (blue dots) are compared to the integral approximation with $\varepsilon=1$ (black line) and the SEM expansion (red line).}
    \label{fig:potentialEnergy}
\end{figure}

In the following example, we consider a $d=2$ model of two Néel Skyrmions in a square lattice of dipolar interacting classical Heisenberg spins, see Fig.~\ref{fig:potentialEnergy}~(a). The Skyrmions have a domain wall width $\lambda = 5$, their core has a radius of $26/5\,\lambda$, and they are separated by a distance $15\,\lambda$, where the parameters for the Skyrmions, as well as their profile, have been taken from \cite{wang2018theory}. We denote the spin orientation at lattice site $\bm y$ as $\bm S(\bm y)$ with $\vert \bm S(\bm y) \vert= 1$. After aligning a central spin $\bm S(\bm x)\to \bm S_c=\bm e_3$ at lattice site $\bm x$, we aim at computing the interaction energy 
\[
U(\bm x)=\bm S_c\cdot \bm H(\bm x)
\] with the surrounding spins. The central spin obeys the equations of motion  \[\frac{\partial\bm S_c}{\partial t}= \bm S_c\times \bm H(\bm x).\] Both equations follow from the effective field $\bm H(\bm x)$, whose continuum representation is given by
\begin{multline*}
 \bm H(\bm x)=\\\frac{1}{V_\Lambda}\ddashint_{\mathds R^2} \frac{\bm S(\bm x+\bm y)\vert \bm y\vert^2-3\bm y(\bm S(\bm x+\bm y)\cdot \bm y)}{\vert\bm y\vert^5}\,\mathrm d \bm y\notag  +\mathcal Z_H(\bm x).
\end{multline*}
Here $\mathcal Z_H(\bm x)$ denotes the lattice contribution that reads in leading order 
\begin{multline*}
 Z_{\Lambda,\nu}\left\vert \begin{matrix}
      \bm 0\\ \bm 0
    \end{matrix}\right\vert \bm S(\bm x)
+3 \Big( (\bm S(\bm x)\cdot \bm \nabla_{\bm y})\bm \nabla_{\bm y}\Big)   Z_{\Lambda,\nu+2}^{\text{reg}}\left\vert \begin{matrix}
      \bm 0\\ \frac{\bm y}{2\pi}
    \end{matrix}\right\vert \Bigg\vert_{\bm y = 0},
\end{multline*}
with $\nu = 3$.
We display the potential energy $U(\bm x)$ for $x_2=0$ as a function of $x_1$ in Fig.~\ref{fig:potentialEnergy}~(b): The blue dots display the energies obtained by exact summation, the black curve shows the standard integral approximation $\mathcal I_\varepsilon(\bm x)$ for $\varepsilon=1$. The red line displays the SEM expansion taking into account derivatives of $\bm S$ up to second order. We observe that the potential energy remains constant both in the center of the two Skyrmions as well as in the far exterior of the domain. Large variations in the potential energy are observed at the boundaries of the Skyrmions. While the integral approximation (black) reproduces the correct qualitative behavior of $U$, it severely fails quantitatively. By including the SEM correction, this significant error is corrected, and the result is visually indistinguishable from the exact value. The SEM approximation thus provides a precision comparable with exact summation but at the numerical cost of the integral approximation. 

\section{Nonlinear systems}
\label{sec:nonlinear_systems}
In the previous section, we have considered, in all generality, systems whose potential energy scales linearly with the function $g$. In many systems of interest, nonlinear effects are of high relevance and need to be considered. In this section, we hence generalize our representation to nonlinear particle interactions. 

\subsection{Derivation of the representation}
\label{subsec:derivation_nonlinear}
We compute the interaction energy $U(\bm x)$ of a test particle at position $\bm x$ with the particles of a distorted lattice with positions $\bm r(\bm x)=\bm x+\bm u(\bm x)$ and the resulting force on the test particle $\bm F(\bm x)$. The nonlinear potential energy of the particle at reference position $\bm x$ due to its interaction with the particles of the lattice reads
\[
U(\bm x)=  \sideset{}{'} \sum_{\bm y \in \Lambda}\vert \bm r(\bm y) - \bm r(\bm x) \vert^{-\nu}.
\]
In order to use the SEM expansion for finding a continuum representation, we first factorize the summand function as $\vert \bm y-\bm x \vert^{-\nu} g(\bm y)$
with 
\[g(\bm y) = \bigg\vert \frac{\bm r(\bm 
y) - \bm r(\bm x)}{\vert \bm y-\bm x\vert }\bigg\vert^{-\nu}. \]
Due to the nonlinearity, the arising function $g$ has an essential singularity at position $\bm x$. However, if we fix an arbitrary direction $\bm y\neq \bm 0$, then $g(\bm x+h\bm y)$ remains smooth in $h\in \mathds R$. Hence, albeit the essential singularity of $g$ at $\bm x$, we can still perform a Taylor expansion in $h$. In close analogy to the derivation of the SEM expansion from the previous section, the SEM operator then reads 
\[
\mathcal D^{(\ell)}g(\bm x)= \sum_{k=0}^{2\ell+1} \frac{1}{k!}\sideset{}{'}\sum_{\bm y \in (\Lambda-\bm x)}\frac{1}{\vert \bm y\vert^\nu} \frac{\partial^k}{\partial h^{k}} g(\bm x+h \bm y)\bigg\vert_{h=0}.
\]
For the lowest order contribution, we find that 
\[
    \frac{1}{\vert \bm y\vert^\nu} g(\bm x+h \bm y)\Big\vert_{h=0}=\vert  D_{\bm y}\bm r(\bm x) \vert^{-\nu },
\]
with the directional derivative $D_{\bm y}=\bm y \cdot \bm \nabla$. 
Thus the continuum representation of the potential energy reads 
\begin{align*}
U(\bm x) &= \frac{1}{V_\Lambda} \ddashint_{\mathds R^d}\vert \bm r( 
\bm y) - \bm r(\bm x) \vert^{-\nu}\,\mathrm d \bm y+ \mathcal Z_U(\bm x),
\end{align*}
with the lowest-order lattice contribution
\begin{equation}
\mathcal Z_U(\bm x)\approx Z_{\Lambda(\bm x),\nu}\left\vert \begin{matrix}
      \bm x\\ \bm 0
    \end{matrix}\right\vert.
    \label{eq:nonlinear_U}
\end{equation}
Here $\Lambda(\bm x) =\bm \nabla \bm r(\bm x)^T \Lambda $ denotes the locally distorted lattice at the position of the test particle $\bm x$,  where $(\bm \nabla \bm r(\bm x))_{ij}=\partial_{x_j} r_i(\bm x)$. 
The corresponding force $\bm F$ on the test particle then follows as
\[
\bm F(\bm x) \approx \frac{1}{V_\Lambda} \ddashint_{\mathds R^d}(-\nu)\frac{ \bm r(\bm y) - \bm r(\bm x)}{\vert \bm r(\bm y) - \bm r(\bm x) \vert^{-(\nu+2)}}\,\mathrm d \bm y\notag \\ + \mathcal Z_F(\bm x),
\]
with the lattice contribution 
\begin{equation}
\mathcal Z_F=\sideset{}{'}\sum_{\bm y\in \Lambda}\frac{1}{2}D_{\bm y} \frac{\partial}{\partial (D_{\bm y} \bm r)}\vert D_{\bm y} \bm r \vert^{-\nu },
\label{eq:nonlinear_F}
\end{equation}
and remaining corrections that scale as fourth derivatives of $\bm r$.
After expanding the summand function as follows
\begin{multline}
    D_{\bm y} \frac{\partial}{\partial (D_{\bm y} \bm r)}\vert D_{\bm y} \bm r \vert^{-\nu }=-\nu D_{\bm y} \frac{D_{\bm y} \bm r}{\vert D_{\bm y} \bm r\vert^{\nu+2}} \\
    =-\nu \frac{D_{\bm y}^2 \bm r}{\vert D_{\bm y} \bm r\vert^{\nu+2}}+\nu(\nu+2)\frac{D_{\bm y}\bm r\Big(D_{\bm y}\bm r\cdot D_{\bm y}^2\bm r\Big)}{\vert D_{\bm y}\bm r \vert^{\nu+4}},
\end{multline}
we see that the resulting Dirichlet series can again be written in terms of higher order derivatives of $\bm r$ where the coefficients are  Epstein zeta functions that include the locally distorted lattice. 

We conclude that, due to the nonlinearity, the effect of the lattice distortion enters in the lattice sums for the lattice contribution. From the scaling argument in Sec.~\ref{subsec:scaling}, we observe that both for the potential energy and for the forces, the lattice contribution becomes particularly relevant in the limit $\nu \to d$, where the pole of the zeta function cancels with the pole of the Hadamard integral.

\subsection{Example 2: Nonlinear Coulomb forces in ion chains}
\label{subsec:ion_chain}

In the following example, we study long-range forces in one-dimensional crystals with long-range interactions, in particular, ion chains. Chains of trapped ions have been a central object of study in the past years, as they are one of the main candidates for qubits in a scalable quantum computer \cite{monroe2013scaling,pogorelov2021compact,olsacher2020scalable,jain2020scalable,wright2019benchmarking,pagano2018cryogenic}; recently an ion trap quantum computer with 21 qubits has been realized \cite{pogorelov2021compact}. Furthermore, ion crystals can be used as quantum simulators for condensed matter systems, for instance, for lattice gauge theories \cite{martinez2016real,manovitz2020quantum}. Long-range interactions between spins of atomic ions can be generated by means of optical dipole forces, where the resulting system can be described by a sine-Gordon model with long-range interactions \cite{maghrebi2017continuous}. When superimposing an additional periodic corrugation potential onto the ion crystal, the resulting system can be used as a quantum simulator for friction on the nanoscale \cite{benassi2011nanofriction,bylinskii2016observation,gangloff2020kinks}. Recently, quantum effects in the associated Aubry transition have been investigated \cite{bonetti2021quantum}. Long-range interactions, either due to the Coulomb repulsion or optically-induced spin-spin interactions, play an important role in ion chains \cite{gambetta2020long,maghrebi2017continuous}. In particular, the correct description of the Coulomb repulsion in a continuum treatment is a challenging task, as the
discreteness of the lattice is relevant at all scales \cite{Dubin1997}. 

We now show how to rigorously include the lattice contribution in the study of the nonlinear long-range forces in a one-dimensional crystal. We analyze the forces that arise in defects in an infinite one-dimensional long-range interacting crystal in a sinusoidal substrate potential $V_{\text{sub}}(r)=\kappa(1- \cos(2\pi  r))$, with $\kappa>0$ the substrate amplitude. In particular, we focus on the case of an ion chain, where the particles interact via the Coulomb repulsion, i.e., $\nu = 1$. The potential energy and the resulting force on the particle at position $x$ due to the long-range interaction then read
\begin{subequations}
\begin{align}
U( x)&=  \sideset{}{'} \sum_{ y \in  \Lambda}\vert  r( y) -  r( x) \vert^{-\nu},\\ 
F(x) &= \sideset{}{'} \sum_{ y \in \Lambda} (-\nu)\frac{  r( 
y) -  r( x)}{\vert  r( 
y) -  r( x) \vert^{-(\nu+2)}}.
\end{align}
\end{subequations}
For $\Lambda = \mathds Z$ and $x\in \Lambda$, the lowest order lattice contributions from Eqs.~\eqref{eq:nonlinear_U}-\eqref{eq:nonlinear_F} take the particularly simple form
\begin{subequations}
\begin{align}
    \mathcal Z_U(x)  &\approx  2\zeta(\nu)s\big(a(x)\big),     \label{eq:finite_size_correction_1d_U}
   \\ 
    \mathcal Z_F(x) &\approx 
    \zeta(\nu)s''(a(x))  r''(x),
    \label{eq:finite_size_correction_1d_F}
\end{align}
\end{subequations}
with the locally modified lattice constant $a(x)=r'(x)$.
 Here $a(x)$ appears in both corrections; in the case of the energy, it appears as an argument in the interaction $s$ and, for the force, as an argument to the elastic constant $K\propto s''$. The result for the lattice contribution obtained from a linearization of the forces is recovered if we set $a(x) = 1$ and neglect the result of the modification of the lattice constant on $s$ and $K$.
 
\begin{figure}
    \centering
    \includegraphics[width=.48\textwidth]{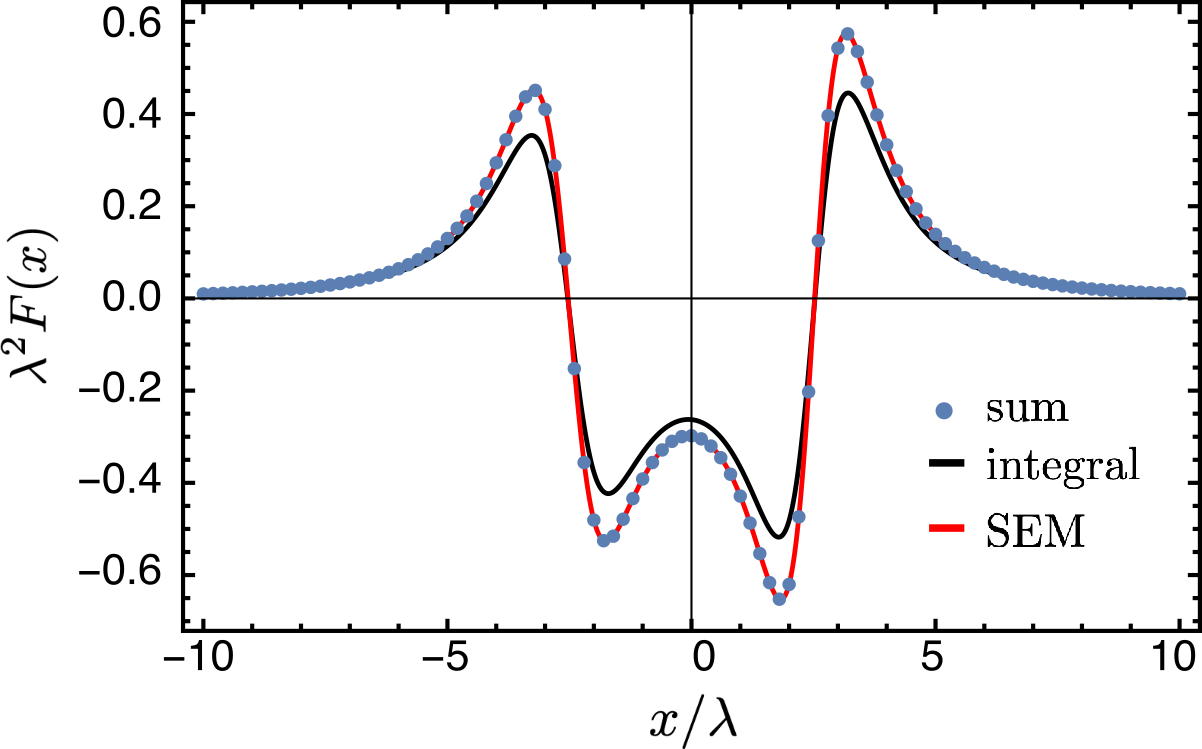}
    \caption{Nonlinear forces in a one-dimensional Coulomb crystal with a breather excitation (bound state of a kink and an anti-kink) for a kink width $\lambda = 5$ and a kink separation of $5\lambda$. The blue dots show the exact forces, the black line displays the continuum approximation with $\varepsilon=1$, and the red line shows the lowest order nonlinear SEM. The asymmetry in the forces is due to the nonlinear interaction. The integral approximation is found to be imprecise, whereas the lowest-order nonlinear SEM yields an excellent approximation.}
    \label{fig:coulomb}
\end{figure}
 The equations of motion in the lowest order SEM expansion (including the substrate potential) then correspond to a sine--Gordon model with nonlinear long-range interactions,
\begin{align}
    \frac{\partial^2  r( x)}{\partial t^2}=\frac{1}{V_\Lambda} \ddashint_{\mathds R}(-\nu)\frac{  r( 
y) -  r( x)}{\vert  r( 
y) -  r( x) \vert^{-(\nu+2)}}\,\mathrm d  y\notag \\+ \zeta(\nu)s''\big(r'(x)\big)  r''(x)+2\pi \kappa\sin(2\pi r(x)),
\end{align}
where we take the limit $\nu \to 1$ to recover the Coulomb interaction.
As the defect, we choose a breather excitation, a bound kink-antikink pair with individual kink widths $\lambda=5$ and a kink-antikink separation $5\lambda$, where the kink profile is modeled via an integral over a normalized Lorentzian. In Fig.~\ref{fig:coulomb}, we display the Coulomb forces (in units of $s''(1)$) obtained from exact summation (blue), the continuum approximation for $\varepsilon=1$, and the lowest order SEM with the lattice contribution in Eq.~\eqref{eq:finite_size_correction_1d_F}. We find that all three computations yield the correct qualitative force behavior. The particles in the chain are drawn towards the kink on the left, as it describes a delocalized particle-hole, whereas the anti-kink on the right describes an excess particle in the chain, from which the remaining particles are repelled. However, the integral approximation severely underestimates the absolute value of the forces. On the other hand, the lowest-order nonlinear SEM correction offers a highly precise approximation to the force sum, which is visually indistinguishable from the exact result and can be efficiently computed.

\section{Multi-atomic lattices}
\label{sec:multi_atomic}
Previously, we have studied lattices with a single atom per unit cell. In the following, we generalize our representation and consider an $n$-atomic and $\Lambda$-periodic lattice $L$,  
\[
 L= \sum_{j=1}^n (\Lambda + \bm d_j),
\]
where  $\bm d_j$, $j=1,\dots, n$ are the positions of the atoms inside the unit cell. Hence, the $n$-atomic lattice $L$ consists of $n$ sublattices, where each may consist of its own atomic species, whose properties are described by different functions $g_j$, $j=1,\dots,n$.

\subsection{Derivation of the representation}
\label{subsec:derivation_multi_atomic}
For simplicity, we focus on linear systems in the following; the nonlinear case can, however, be treated in close analogy.
We consider the interaction energy $U(\bm x)$ of a test particle at position $\bm x$ with the multi-atomic lattice $L$,
\[
    U(\bm x) = \sum_{j=1}^n\,\sideset{}{'}\sum_{\bm y\in (\Lambda + \bm d_j)} \frac{g_j(\bm y)}{\vert \bm y-\bm x \vert^\nu}.
\]
In case that $\bm x \in L$, the corresponding self-energy term is excluded. Now, we apply the SEM expansion in Eq.~\eqref{eq:sem_expansion} for the mono-atomic lattice $\Lambda$, such that
\begin{equation*}
U(\bm x) = \sum_{j=1}^n \Bigg( \frac{1}{V_\Lambda} \ddashint \limits_{\mathds R^d} \frac{g_j(\bm y)}{\vert \bm y-\bm x\vert^\nu}\, \mathrm d \bm y+Z_{\Lambda,\nu}^{\text{reg}}\left\vert \begin{matrix}
      \bm x-\bm d_j\\\frac{i\bm \nabla}{2\pi }
    \end{matrix}\right \vert \,g_j(\bm x)\Bigg)
\end{equation*}
follows as the direct generalization of the corresponding $U(\bm x)$ in the mono-atomic case. We point out that the continuum approximation does not include the positions of the particles in the unit cell. It rather describes the interaction of a test particle immersed in $n$ different continua with different properties. The discrete structure of the lattice, the positions of the atoms in the unit cell $\bm d_j$, and, in particular, the distance of the test particle to its nearest neighbor, are fully encoded in the first argument of the Epstein zeta function that describes the lattice contribution.

A particularly simple case arises when all the particles in the lattice are identical and hence $g_j=g$. Then,
\begin{equation}
    U(\bm x) =  \frac{n}{V_\Lambda} \ddashint \limits_{\mathds R^d} \frac{g(\bm y)}{\vert \bm y-\bm x\vert^\nu}\, \mathrm d \bm y+\sum_{j=1}^n Z_{\Lambda,\nu}^{\text{reg}}\left\vert \begin{matrix}
      \bm x-\bm d_j\\\frac{i\bm \nabla}{2\pi }
    \end{matrix}\right \vert g(\bm x).
    \label{eq:SEM_multi_atomic}
\end{equation}
Here, only a single integral needs to be computed, and the $n$ differential operators reduce to a single differential operator acting on $g$. This situation appears, among others, in the case of carbon atoms in diamond.

\subsection{Example 3: Spin-wave in a 3D Pyrochlore lattice}
\label{subsec:multi_atomic_example}

\begin{figure}
    \centering
    \includegraphics[width=.5\textwidth]{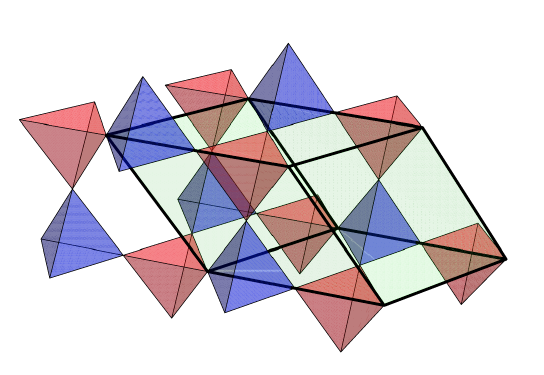}
    \caption{Pyrochlore lattice with $n=4$ atoms per elementary lattice cell (green parallelepiped). The atoms are placed at the corners of the tetrahedra.}
    \label{fig:pyLattice}
\end{figure}
As an example of the relevance of the lattice contribution in a multi-atomic system, we now consider a spin-wave in a three-dimensional Heisenberg spin lattice with dipolar long-range interactions. The identical spins shall be arranged in the pyrochlore crystal structure that exhibits $n=4$ atoms per unit cell. The lattice can be understood in terms of corner-sharing tetrahedra, see Fig.~\ref{fig:pyLattice}, where each corner is occupied by a particle.  Details on the crystal structure are given in Appendix~\ref{sec:app_pyrochlore}.

More than 20 years ago, spin ice, a magnetic analog to water ice, was found in ferromagnetic pyrochlore materials \cite{PhysRevLett.79.2554, Ramirez1999}. These systems are well described by classical spins with strong Ising anisotropy \cite{PhysRevB.79.014408,bramwell2001spin} but with long-range dipolar interactions, which play an important role in the origin of the spin ice formation \cite{PhysRevLett.84.3430, PhysRevLett.83.1854}.
The discovery of magnetic monopoles has brought spin ice to the attention of a wide community \cite{Castelnovo2008,Bramwell2009,doi:10.1126/science.1178868,Jaubert2009,ross2011quantum}, in which the dipole-dipole interaction translates into an effective Coulomb interaction between magnetic monopoles \cite{Bramwell_2020}. This has also been the starting point for many more investigations in such emergent systems, such as artificial spin ice \cite{Skj2020}, quantum spin ice \cite{Gingras_2014}, monopole shot noise \cite{Dusad2019}, as well as recently engineering emergent quantum electrodynamics \cite{PhysRevLett.127.117205} in spin ice materials.

\begin{figure}
    \centering
    \includegraphics[width=.47\textwidth]{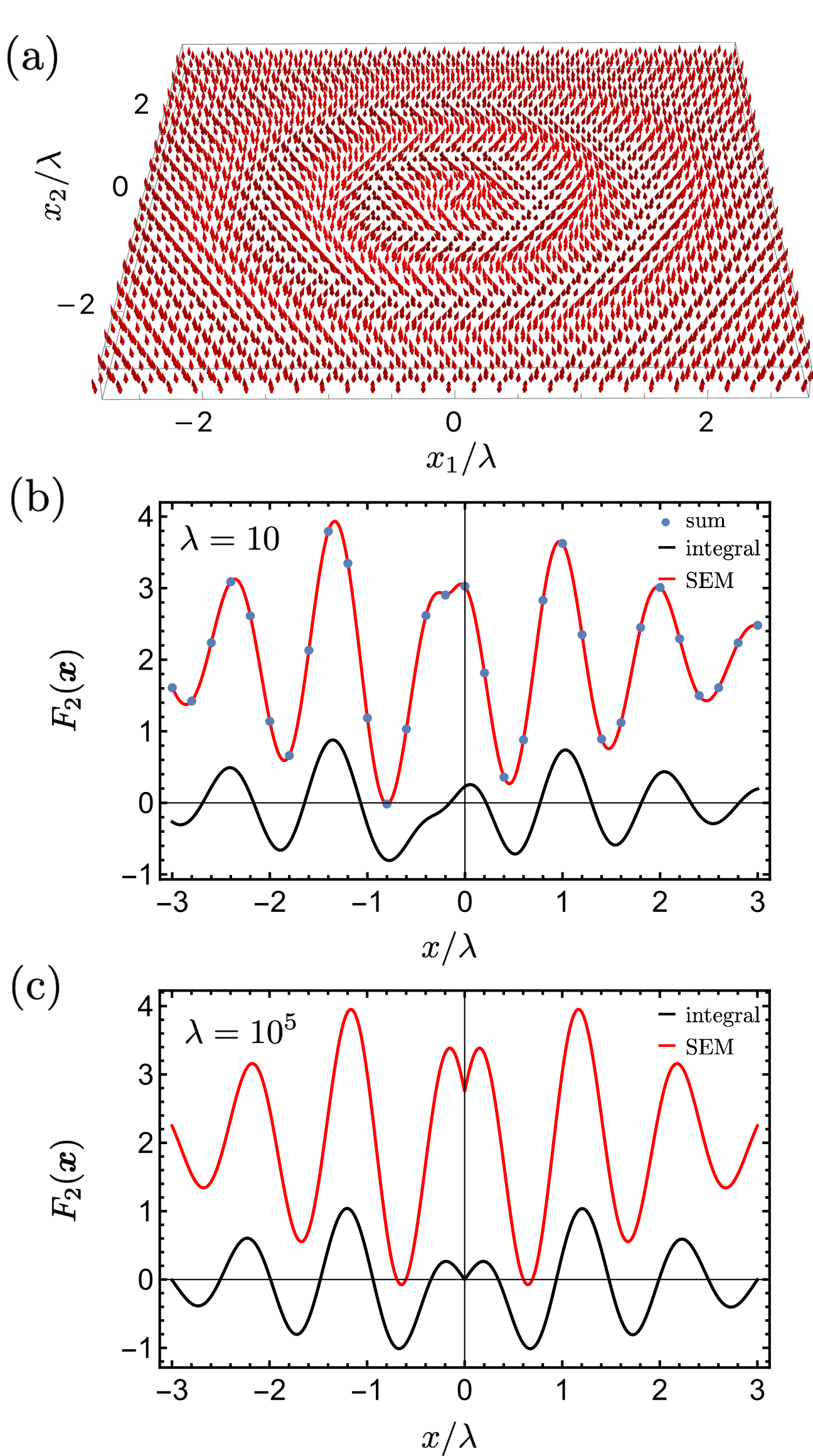}
    \caption{(a) Spherical spin wave with origin at $\bm x=0$, wavelength $\lambda=10$ and amplitude $\theta = 3/10$ in in a three-dimensional pyrochlore lattice; slice through one elementary lattice cell around $x_3=0$. (b) Dipole force along $\bm e_2$ on a test spin  $\bm S_c=\bm e_3$ positioned at $\bm x_0+x\bm a_1/\vert \bm a_1\vert$ with the position in the elementary lattice cell $\bm x_0=(\bm a_1+\bm a_2+\bm a_3)/2$ and the lattice vector $\bm a_1=-(1,\sqrt{3},0)$ for $\lambda = 10$. The SEM expansion (red), including up to fourth order derivatives, faithfully reproduces the sum (blue), whereas the integral approximation (black) with $\varepsilon=\vert \bm a_1 \vert$ fails. (c) Dipole force as in (b) for a macroscopic wavelength $\lambda = 10^5$ where exact summation becomes impossible. The results in (c) are symmetric around $x=0$ in contrast to (b) due to $\vert \bm x_0\vert\ll \lambda$.}
    \label{fig:pySpinWave}
\end{figure}

In the following, we focus on the classical model and study the full long-range ferromagnetic dipole interaction. We compute the forces that are exerted by a spin wave on a test spin. We consider a spherical spin wave centered at $\bm x=0$ with angular momentum axis $\bm e_3$, wavelength $\lambda$, and amplitude $\theta$. The spin vector $\bm S(\bm x)$ can be modeled as
\begin{equation*}
 \bm S(\bm x) =
 \left(\begin{matrix}\sin(\theta(\bm x)) \cos(2\pi \vert \bm x\vert/\lambda)  \\
 \sin(\theta(\bm x)) \sin(2\pi \vert \bm x\vert/\lambda)\\ \cos(\theta(\bm x))
 \end{matrix}\right),
\end{equation*}
where the amplitude $\theta$ shall decay as 
\[
    \theta(\bm x) = \theta_0 e^{-\vert \bm x \vert^2/\gamma^2}.
\] 
For our study, we make the parameter choice $\lambda = 10$, $\gamma = 2.5\,\lambda$ and $\theta_0=3/10$. The spin wave is displayed in Fig.~\ref{fig:pySpinWave}, where we show a slice of the pyrochlore lattice around $x_3=0$. We now determine the interaction of the full lattice with a test spin $\bm S_c = \bm e_3$ positioned at $\bm x$.
The total dipole force reads
\[
\bm F(\bm x)= \sum_{j=1}^n\,\sideset{}{'}\sum_{\bm y\in (\Lambda + \bm d_j-\bm x)} \frac{g(\bm y+\bm x)}{\vert \bm y\vert^5},
\]
where 
\[
g(\bm y+\bm x) = \bm S_c \times \Big(\bm S(\bm y+\bm x)\vert \bm y\vert^2 - 3  \bm y(\bm S(\bm y +\bm x)\cdot \bm y)\Big).
\]
We now position the spin at
\[
    \bm x = \bm x_0 + x \frac{ \bm a_1}{\vert \bm a_1\vert},
\]
with $x/\vert \bm a_1 \vert$ an integer, such that $\bm x_0$ describes the position of the test spin in the elementary lattice cell.
In the following, we position the test particle in the center of the elementary lattice cell $\bm x_0=(\bm a_1+\bm a_2+\bm a_3)/2$.
We now approximate the dipole interaction of the lattice with the test particle by means of the SEM expansion from Eq.~\eqref{eq:SEM_multi_atomic}.
Note that here only $\bm x_0$ enters as an argument to the Epstein zeta function due to $\Lambda$-periodicity. 
We display the $\bm e_2$-component of the force as a function of $x$ in Fig.~\ref{fig:pySpinWave} (b).
The exact forces (blue) are precisely reproduced by the SEM expansion (red).
The integral approximation (black) with $\varepsilon = \vert \bm a_1\vert=2$, however, fails in describing the quantitative and qualitative behavior.
The same holds true if we increase the wavelength to the macroscopic value $\lambda = 10^5$ in (c).
In this case,  exact summation is not available anymore as the required summation task becomes impossible, even on specialized hardware.
The scaling of the remainder of the SEM expansion, however, guarantees that the expansion error falls off polynomially as $\lambda$ increases, such that, in particular for large $\lambda$, the SEM result is equivalent to exact summation for all practical purposes.
The force resembles the rescaled result for the smaller wavelength in (b), where, however, now the result is symmetric in $x$ as $\vert \bm x_0\vert\ll \lambda$.
Hence, the SEM expansion offers a powerful tool for describing the long-range dipole interaction in three-dimensional lattices.

\textcolor{myred}{In Appendix \ref{sec:anomalous_dispersion}, we show that our representation also allows for the computation of the analytic quantum spin wave dispersion relation in multi-atomic lattices in arbitrary dimensions. Here, we demonstrate that, depending on the system dimension and the exponent of the interaction, anomalous dispersion can occur, where the spin wave energy in the large wavelength limit does not obey a $\vert \bm k\vert^2$ scaling anymore, which significantly changes the behavior of the system and generalizes recent results in one and two dimensions~\cite{yusuf2004spin,peter2012anomalous,PhysRevLett.111.207202}.}

\color{myred}

\section{Unconventional superconductivity in long-range interacting systems}

\label{sec:unconventional_superconductivity}

Superconductivity is a key macroscopic quantum phenomenon in condensed matter physics. Recently, long-range interactions have become a focus in the field of superconductivity, as they can play an important role in various different systems.

In angle-resolved photoemission experiments on the 1D cuprate material Ba$_{2-x}$Sr$_x$CuO$_{3+\delta}$, spectroscopic signatures indicate a non-local superconducting glue \cite{doi:10.1126/science.abf5174}. Phonon-mediated long-range interactions may play a critical role in understanding the pairing mechanism in these unconventional superconductors beyond pure electronic correlations \cite{PhysRevLett.127.197003}.  

In addition to their potential role in high-temperature superconductivity, long-range interactions can also be induced in THz nanoplasmonic cavities, in which the electrons couple strongly to external laser photons, inducing long-range density-density interactions that can be controlled by tuning the laser field \cite{PhysRevLett.125.053602,2021NatCo12.5901C}.

Effective long-range interactions can also emerge in superconducting heterostructures, such as magnetic nanowires and islands. By carefully designing these structures, it is possible to create new states of matter that exhibit unique properties such as topological superconductivity \cite{PhysRevLett.120.017001}. They might also have a strong effect on Majorana zero modes, relevant for topological quantum computing platforms \cite{PhysRevB.88.155420, PhysRevLett.113.156402}. 

So far, most theoretical work on superconductivity has focused on short-range interactions, with the BCS theory being the most widely accepted approach for explaining superconductivity in various materials. Given the potential role of long-range interactions, a theoretical description taking these interactions into account is necessary to understand the underlying effects and the mechanisms of superconductivity in these systems.

In this section, we use our continuum representation to derive a generalized BCS theory applicable to algebraically decaying long-ranged electron-electron interactions. We derive a generalized gap equation valid for singlet and triplet pairing and also the time evolution of the superconducting state to describe non-equilibrium Higgs oscillations. 

We apply this approach to calculate the phase diagram of a 2D BCS superconductor with long-range interactions as a function of the long-range interaction strength, the on-site interaction strength, and the decay exponent. We find a rich phase diagram with unconventional s-, p- and d-wave pairings and quantum-critical, continuous phase transitions. 

We also calculate non-equilibrium Higgs oscillations in various phases, demonstrating the powerfulness of our approach for non-equilibrium dynamics.
The Higgs mode is a solid-state analog to the famous Higgs particle and has been measured experimentally using intense terahertz pulses in conventional s-wave superconductors  \cite{PhysRevLett.109.187002, PhysRevLett.111.057002, doi:10.1126/science.1254697}. Recent experimental progress focused on Higgs modes in unconventional superconductors and multi-band superconductors  \cite{vaswani2021light, PhysRevLett.120.117001,2020NatCo11.1793C,doi:10.1146/annurev-conmatphys-031119-050813} as well as Higgs modes in light-induced superconductors \cite{Luo2022, PhysRevB.101.180507} with various theoretical predictions concerning classification as well as non-equilibrium signals of Higgs modes \cite{doi:10.1146/annurev-conmatphys-031214-014350,PhysRevB.76.224522,PhysRevB.78.132505,PhysRevB.77.180509,PhysRevB.88.104511,2013EL10117002A,PhysRevB.92.224517,PhysRevB.96.184518,PhysRevB.95.104503,PhysRevB.100.140501,2016NatCo711921K,schwarz2020classification,PhysRevB.101.224510,PhysRevB.104.174508}.

\subsection{Model introduction and continuum representation in Fourier space}
We assume that the electrons are tightly bound to the ions on a lattice $\Lambda$ with a Hamiltonian
\[
H= H_0+H_\text{int}.
\]
Here $H_0$ describes hopping between neighboring lattice sites at rate $\tau>0$ ($\hbar =1$),
\[
H_0 = -\frac{\tau}{2} \sum_{\sigma} \sum_{\bm x\in \Lambda} \sum_{i=1}^d \big(c_{\sigma,\bm x}^\dagger c_{\sigma,\bm x+A_\Lambda \bm e_i}+\mathrm{h.c.}\big),
\]
with $c^\dagger_{\sigma,
\bm x}$ and $c_{\sigma,\bm x}$ the creation and annihilation operators for an electron with spin $\sigma\in\{\uparrow,\downarrow\}$ at lattice site $\bm x$.
Additionally, we consider an isotropic electron-electron interaction,
\begin{align*}
H_\text{int}=\frac{1}{2} \sum_{\sigma,\sigma'}  \sum_{\bm x,\bm y \in \Lambda} c^\dagger_{\sigma,\bm x}c^\dagger_{\sigma',\bm x-\bm y} V_{\sigma\sigma'}(\bm y)c_{\sigma',\bm x-\bm y} c_{\sigma,\bm x},
\end{align*}
with an on-site interaction 
\begin{align*}
V_{\sigma\sigma'}(\bm 0) = - C_{\sigma\sigma'}\le 0,
\end{align*}
where $C_{\sigma\sigma'}\propto \delta_{\sigma,-\sigma'}$,
and a power-law long-range interaction
\begin{align*}
V_{\sigma\sigma'}(\bm y) =- U_{\sigma\sigma'} \frac{1}{\vert \bm y \vert^\nu} \le 0,\quad \bm y \neq 0,
\end{align*}
with constants $C_{\sigma\sigma'},U_{\sigma\sigma'} \geq 0$ corresponding to an attractive potential.
We then proceed by writing the operators in Fourier space 
\[
c_{\sigma,\bm x}=\sqrt{V_\Lambda} \int \limits_{E^\ast } e^{2\pi i \bm k\cdot \bm x}c_{\sigma}(\bm k)\,\mathrm d \bm k,
\]
with $E^\ast$ the first Brillouin zone. 
The Fourier transform diagonalizes the quadratic hopping Hamiltonian,
\[
H_0 = \sum_{\sigma}\int \limits_{E^\ast} \xi(\bm k)c_{\sigma}^\dagger(\bm k) c_{\sigma}(\bm k)\,\mathrm d \bm k,
\]
with the electron dispersion relation in the normal state
\[
\xi(\bm k) = -\tau \sum_{i=1}^d \cos(\bm k\cdot A_{\Lambda}\bm e_i).
\]
The singular interaction in real space is transformed into an integral over the Epstein zeta function in Fourier space, namely 
\begin{align*}
H_\text{int}&=-\frac{V_\Lambda}{2} \sum_{\sigma,\sigma'} \int \limits_{E^\ast} \int\limits_{E^\ast }\ddashint \limits_{E^\ast} \Big(C_{\sigma\sigma'} +U_{\sigma\sigma'}Z_{\Lambda,\nu} \left \vert \begin{matrix} 0 \\ \bm q \end{matrix} \right \vert\Big) \\ & \times  c_{\sigma}^\dagger (\bm k+\bm q) c_{\sigma'}^\dagger (\bm k'-\bm q) c_{\sigma'}(\bm k') c_{\sigma}(\bm k) \,\mathrm d \bm q\,\mathrm d \bm k\,\mathrm d \bm k'.
\end{align*}

We now apply the following mean field approximation, replacing the product of the four operators above by
\begin{align*}
& \Big(c_{\sigma}^\dagger (\bm k+\bm q) c_{\sigma'}^\dagger (\bm k'-\bm q) - \Big\langle c_{\sigma}^\dagger (\bm k+\bm q) c_{\sigma'}^\dagger (\bm k'-\bm q) \Big\rangle \Big) \notag \\
\times &\Big(c_{\sigma'}(\bm k') c_{\sigma}(\bm k)-\Big\langle c_{\sigma'}(\bm k') c_{\sigma}(\bm k) \Big\rangle\Big)\approx 0,
\end{align*}
where $\langle A \rangle=\langle \psi \vert A \vert \psi \rangle $ denotes the expectation value of an operator $A$ in the ground state $\vert \psi \rangle$. In the following, we assume that the state  has zero momentum, 
\[
 \big\langle c_{\sigma'}(\bm k') c_{\sigma}(\bm k)\rangle    \approx \delta(\bm k+\bm k')  \alpha_{\sigma\sigma'}(\bm k)
\]
with the correlation matrix $\alpha$ given by \[
 \alpha_{\sigma\sigma'}(\bm k) =\int \limits_{E^\ast} \langle c_{\sigma'}(\bm k') c_{\sigma}(\bm k)\rangle   \,\mathrm d \bm k'.
\]
Noting that $c_{\sigma,\bm k}^\dagger$ is $\Lambda^*$-periodic in $\bm k$, we can make the substitution $\bm k
\to \bm k-\bm q$. This allows us to rewrite the full Hamiltonian as
\begin{align*}
H&= \sum_{ \sigma} \int \limits_{E^\ast} \xi(\bm k) c^\dagger_{\sigma} (\bm k) c_\sigma(\bm k) \,\mathrm d\bm k \notag \\&- \frac{1}{2}\sum_{\sigma,\sigma'} \int \limits_{E^\ast}   \big(\Delta_{\sigma\sigma'}(\bm k) c_{\sigma}^\dagger(\bm k) c_{\sigma'}^\dagger(-\bm k)+\text{h.c.} \big) \,\mathrm d \bm k .
\end{align*}
Here, $\Delta(\bm k)\in \mathds C^{2\times 2}$ is the superconducting gap matrix,
\[
\Delta_{\sigma\sigma'}(\bm k)= V_\Lambda\ddashint \limits_{E^\ast}  \bigg(C_{\sigma\sigma'}+U_{\sigma\sigma'} Z_{\Lambda,\nu} \left \vert \begin{matrix} 0 \\ \bm q \end{matrix} \right \vert \bigg) \alpha_{\sigma\sigma'}(\bm k -\bm q)\,\mathrm d \bm q.
\]
Note that the gap matrix depends on the correlation matrix $\alpha$, which is determined by the current state of the superconductor.

Any simple discretization of the above integral is bound to lead to errors on a qualitative level due to the singularity of the Epstein zeta function at $\bm q=0$. This problem is overcome by using the continuum representation in Fourier space. To this end, the zeta function is separated into the singular part at $\bm q=0$ and the well-behaved regularized Epstein zeta function, \[
Z_{\Lambda,\nu} \left \vert \begin{matrix} 0 \\ \bm q \end{matrix} \right \vert= \frac{\hat s_\nu(\bm q)}{V_\Lambda}+Z^{\text{reg}}_{\Lambda,\nu} \left \vert \begin{matrix} 0 \\ \bm q \end{matrix} \right \vert
\]
with $\hat s_\nu(\bm q)\propto \vert \bm q\vert^{\nu-d}$ the 
Fourier transform on $\mathds R^d$ of the interaction, see Eq.~\eqref{eq:fourier-interaction}. The splitting in continuum and lattice contributions then reads
\begin{equation*}
\Delta(\bm k) = \mathcal I_\Delta(\bm k) + \mathcal Z_\Delta(\bm k),
\end{equation*}
with
\begin{align*}
\mathcal I_{\Delta}(\bm k)&= \ddashint \limits_{E^\ast}  \bigg(C_{\sigma\sigma'}+U_{\sigma\sigma'} \hat s_\nu(\bm q)\bigg) \alpha_{\sigma\sigma'}(\bm k -\bm q)\,\mathrm d \bm q,\notag \\
\mathcal Z_{\Delta}(\bm k) &= V_\Lambda\ddashint \limits_{E^\ast}  \bigg(C_{\sigma\sigma'}+U_{\sigma\sigma'} Z_{\Lambda,\nu}^{\text{reg}} \left \vert \begin{matrix} 0 \\ \bm q \end{matrix} \right \vert \bigg) \alpha_{\sigma\sigma'}(\bm k -\bm q)\,\mathrm d \bm q.
\end{align*}
Subsequently, the smooth regularised Epstein zeta function can be expanded  in an absolutely convergent Taylor series in $\bm q$ around $\bm 0$, corresponding to the expansion of the lattice contribution in terms of higher-order derivatives in real space. Fourier space, however, offers the advantage that all orders can be included simultaneously. A special quadrature must be applied to integrate the singularity in the continuum contribution faithfully. This can, for instance, be achieved through a Duffy transformation. More details on computational aspects are given in Appendix~\ref{sec:numerics}.

\subsection{Generalized long-range BCS gap equation}
\label{sec:BCS_ground_state}
We now determine the spectrum of $\mathcal H(\bm k)$, diagonalize the Hamiltonian, and determine the equation for the stationary superconducting gap matrix. We first define an auxiliary vector $\bm \Psi$ that includes the fermionic operators,
\[
\bm \Psi (\bm k) = \Big(c_{\uparrow}(\bm k), c_{\downarrow}(\bm k), c^\dagger_{\uparrow}(-\bm k),c^\dagger_{\downarrow}(-\bm k)\Big)^T.
\]
After removing a constant term and using that $\xi(-\bm k)=\xi(\bm k)$,  we can cast the Hamiltonian in the form 
\[
    H= \int \limits_{E^*} \frac{1}{2} \bm \Psi^\dagger (\bm k) \mathcal H(\bm k) \bm \Psi(\bm k)\,\mathrm d \bm k,
\]
with
\[
\mathcal H(\bm k)=\left( 
    \begin{matrix}
     \xi(\bm k) \mathds 1_2 & -\Delta(\bm k) \\ -\Delta^\dagger (\bm k)  & -\xi(\bm k) \mathds 1_2
    \end{matrix}
    \right),\quad
 \Delta = \left( \begin{matrix}
    \Delta_{\uparrow\uparrow} & \Delta_{\uparrow\downarrow} \\ \Delta_{\downarrow\uparrow}  & \Delta_{\downarrow\downarrow} 
    \end{matrix} \right).
\]
The energy spectrum of the Hamiltonian includes four branches $\pm E_1(\bm k)$ and $\pm E_2(\bm k)$, where $E_1$ and $E_2$ are the eigenvalues of the matrix operator
\[
E[\Delta] =\sqrt{\xi^2 \mathds 1_2 +\Delta^\dagger \Delta}.
\]

We subsequently determine the ground state of the superconductor. To this end, we diagonalize the Hamiltonian through a standard Bogoliubov transformation and determine the ground state, details of which are given in Appendix~\ref{sec:bogoliubov_transform}.  
The ground state is then fully determined by the density matrix $\rho$,
\[
\rho_{i,j}(\bm k) = \frac{1}{2} \ddashint_{E^*} \langle \psi_\text{BCS}\vert \Psi^\dagger_j(\bm k') \Psi_i(\bm k) \vert \psi_\text{BCS}\rangle \,\mathrm d \bm k',
\]
that includes all two-body correlations. In particular, we can identify $\alpha(\bm k)/2$ as the $2\times2$ block matrix on its upper right. The density matrix in the ground state reads
\[
\rho= \frac{1}{4}\left(
\begin{matrix} 
\mathds 1_2 -\xi E[\Delta^\dagger]^{-1} & \Delta E[\Delta]^{-1}\\
\Delta^\dagger E[\Delta^\dagger]^{-1} & \mathds 1_2 + \xi E[\Delta]^{-1}
\end{matrix}\right).
\]
and hence we find for the correlation matrix $\alpha$ that
\[
\alpha[\Delta] = \frac{1}{2} \Delta E[\Delta]^{-1},
\]
with the symmetry constraint
\[
\alpha[\Delta](\bm k)=-\alpha[\Delta]^T(-\bm k)
\]
that is imposed by the fermionic anticommutation relations. 
We arrive at the generalized Bardeen--Cooper--Schrieffer (BCS) gap equation,
\begin{equation*}
\Delta_{\sigma\sigma'}(\bm k) = V_\Lambda \ddashint \limits_{E^\ast}  \Big(C_{\sigma\sigma'}+ U_{\sigma\sigma'} Z_{\Lambda,\nu} \left \vert \begin{matrix} 0 \\ \bm q \end{matrix} \right \vert \Big)\alpha_{\sigma\sigma'}[\Delta](\bm k-\bm q)\mathrm d \bm q,
\end{equation*}
that is valid for any lattice,  any power-law interaction, and any space dimension.
In the following, we focus on spin-independent interactions $U_{\sigma\sigma'}=U_0\ge 0$ and $C_{\sigma\sigma'}=C_0\ge 0$, where the generalized gap equation takes the form
\begin{equation}
\Delta(\bm k) = V_\Lambda \ddashint \limits_{E^\ast}  \Big(C_{0}+ U_{0} Z_{\Lambda,\nu} \left \vert \begin{matrix} 0 \\ \bm q \end{matrix} \right \vert \Big)\Big(\frac{1}{2}\Delta E[\Delta]^{-1}\Big)(\bm k-\bm q)\mathrm d \bm q.
\label{eq:generalized_gap_equation}
\end{equation}

Among all solutions to the above gap equation, the solution associated with the lowest energy forms the ground state. This energy is found by collecting all constant energy contributions in the above derivation.
After inserting the stationary gap equation above, the energy follows compactly  as
\begin{equation}
    E_\text{GS}
=  \frac{N}{2} \int \limits_{E^\ast}  \mathrm{Tr} \bigg(\frac{1}{2}\Delta E[\Delta]^{-1} \Delta^\dagger+ \xi \mathds 1_2 -E[\Delta] \bigg)(\bm k)\,\mathrm d \bm k,
\label{eq:ground_state_energy}
\end{equation}
with $N$ an extensive constant proportional to the number of sites.

\subsection{Non-equilibrium dynamics}
We now determine the time evolution of the operators $\bm \Psi$ in the  Heisenberg picture. The Heisenberg equations of motion for the fermionic operators are given by
\[
i \frac{\partial}{\partial t} \bm \Psi_t(\bm k) =[\bm \Psi_t(\bm k),H_t]=\mathcal H_t(\bm k) \bm \Psi_t(\bm k).
\]
As the Hamiltonian is bilinear in the creation and annihilation operators, we can write $
\bm \Psi_t = U_t \bm \Psi_0,$
with the time evolution operator $U_t(\bm k)\in \mathds C^{4\times 4}$
and $U_0(\bm k)=\mathds 1$  the identity matrix. This implies for the time evolution of the density matrix that 
\[
\rho_t = U_t \rho_0 U_t^\dagger, 
\]
where $\alpha_t/2$ is  the block on the upper right. 
The superconducting gap at time $t$ follows from the correlation matrix as  
\[
\Delta_{t}(\bm k) = V_\Lambda \ddashint \limits_{E^\ast}  \Big(C_{0}+ U_{0} Z_{\Lambda,\nu} \left \vert \begin{matrix} 0 \\ \bm q \end{matrix} \right \vert \Big) \alpha_{t}(\bm k-\bm q)\mathrm d \bm q.
\]
Finally, the time evolution operator itself obeys the nonlinear differential equation
\[
i \frac{\partial}{\partial t} U_t = \mathcal H[\Delta_t] U_t.
\]
The numerical solution of the generalized BCS gap equation and the time evolution of the condensate is discussed in Appendix~\ref{sec:numerics}.

\subsection{Pairing structure}
In triplet systems, it is convenient to represent the gap matrix $\Delta$ in the form introduced
by Balian and Werthamer~\cite{balian1963superconductivity,sigrist1991phenomenological},
\[
\Delta =  \left(\begin{matrix} 
-d_x +i d_y & d_z + \psi \\
d_z-\psi & d_x + i d_y
\end{matrix}\right),
\]
with $\bm d=(d_x,d_y,d_z)^T$ an odd function that describes the triplet channel and $\psi$ an even function associated with the singlet channel. A numerical analysis of the solutions in \eqref{eq:generalized_gap_equation} and their energies in \eqref{eq:ground_state_energy} shows that  the gap matrix in the ground state can be brought in the form,
\[
\psi \in \mathds R,\quad i \bm d \in \mathds R^3.
\]
Hence $\bm d$ is purely imaginary if $\psi$ is chosen as real (which is possible due to $\mathrm{U}(1)$ symmetry of the gap function).

Under these conditions, the energy bands degenerate,
\[
E_i=\sqrt{\xi^2+ \vert\psi\vert^2 +\vert \bm d \vert^2 }, \quad i=1,2.
\]
Furthermore, energy and gap equations are invariant under real orthogonal transformations of $\bm d$. For solutions with finite $\bm d$ we find that the pairing in the ground state is always such that 
\[ 
d_y(k_1,k_2)=-d_x(k_2,k_1),\quad d_z=0,
\]
after a suitable orthogonal transformation.
Regarding the symmetry properties of $\bm d$ we have that
\[
d_x(k_1,k_2) \approx i \sin(2\pi k_1)
\]
and thus $\Delta^\dagger \Delta$ is proportional to the identity matrix.
The associated pairing is called unitary \cite{sigrist1991phenomenological}.

\subsection{Emergence of exotic long-range induced  superconducting phases}

\begin{figure*}
\centering
\includegraphics[width=.98\textwidth]{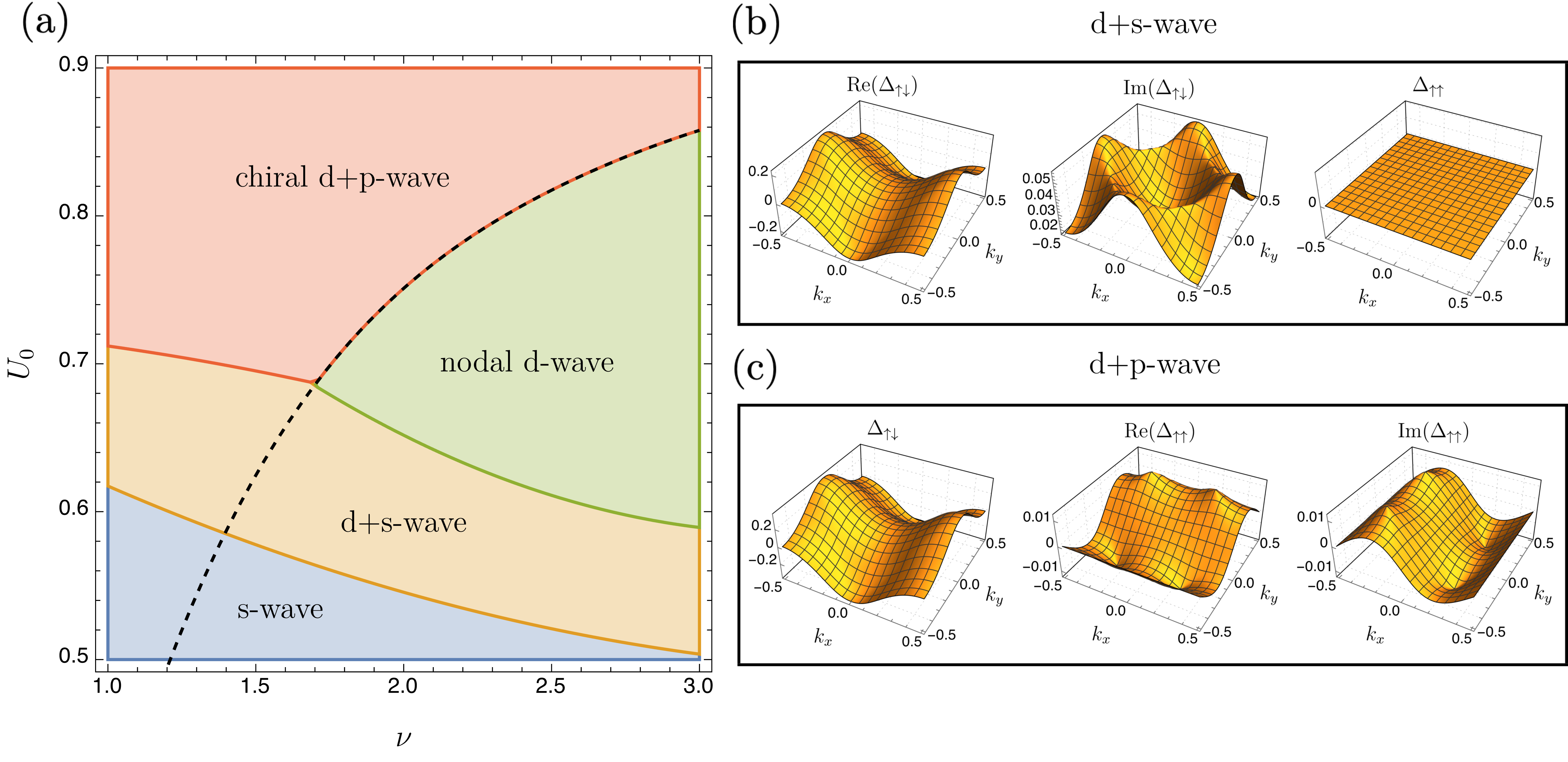}
\caption{\textcolor{myred}{(a) Zero temperature phase diagram for a 2D superconductor, including both singlet and triplet pairing, with power-law electron-electron interactions and lattice structure $\Lambda=\mathds Z^2$ as a function of the interaction exponent $\nu$ and the interaction strength $U_0$ for a fixed onsite interaction $C_0 = 0.75$. The dashed line represents the phase boundary for $C_0= 0$ where only the d-wave and chiral d+p-wave phases exist. All transitions are continuous up to the transition between the phases d+s and d+p, which is of first order. (b) Gap matrix in the d+s phase for $\nu=2.01$ and $U_0=0.6$. (c) Gap matrix in the chiral d+p phase for $\nu=2.01$ and $U_0=0.8$.}}
\label{fig:exotic_phases}
\end{figure*}

We now determine the phase diagram of a 2D superconductor on a square lattice $\Lambda=\mathds Z^2$ with power-law long-range interactions. In the following, we assume a spin-independent  interaction strength
$U_{\sigma\sigma'}=U_0\ge 0$ and same for the onsite-interaction strength $C_{\sigma\sigma'}=C_0\ge 0$.
We vary the power law's interaction exponent $\nu$ and $U_0$ for fixed values of the onsite interaction $C_0$.
The arising phases and their transitions are displayed in Fig.~\ref{fig:exotic_phases}.

We first discuss the case of vanishing on-site interaction $C_0=0$, where only two phases (green and red) separated by the dashed line occur. For the well-studied case of nearest-neighbor interactions, $\nu \to \infty$, the ground state exhibits d-wave symmetry (green, below dashed line, $\psi(\bm k)\sim \cos(2\pi k_1)-\cos(2\pi k_2)$). In addition, the two degenerate energy bands $E_i(\bm k)$ exhibit nodes along the Fermi surface that are energetically costly yet are unavoidable due to the symmetry prescribed by the d-wave solution.

The situation changes significantly when long-range interactions are present. A quantum-critical transition (dashed line) to a new phase (red, above dashed line) occurs for long-range interaction with finite $\nu$.  To avoid the energetically costly band nodes, a p-wave solution in the triplet channel accompanies the d-wave solution in the singlet channel. The exotic phase emerges purely due to the long-range interaction and is topologically nontrivial. In the chiral phase, two degenerate bands with Chern numbers $\pm 1$ appear, where the degeneracy can be lifted by applying an arbitrarily small magnetic field.

Two additional phases emerge in the global phase diagram if a non-zero on-site interaction is added ($C_0 = 0.75$). For vanishing $U_0$, the case of pure on-site interaction, the superconductor assumes standard s-wave symmetry (blue, $\psi \sim\,\text{const}$). As $U_0$ increases, a continuous transition to a mixed state (yellow) occurs where a d-wave solution builds up in the imaginary part of $\psi$ in addition to the real-valued $s$-wave solution (up to $\text{U}(1)$ symmetry). This solution again avoids the nodes of the pure d-wave symmetry and adapts itself to the long-range interaction. As the strength of the long-range interaction increases, we find that the s-wave contribution falls off while the d-wave contribution builds up. Depending on the interaction exponent $\nu$, two different transitions may take place: If $\nu$ is sufficiently small, a first-order transition to the chiral d+p-wave phase (red) occurs. On the other hand, if $\nu$ is sufficiently large, a continuous transition to the nodal d-wave phase (green) takes place.  Note that the discontinuous transition between the d+s to d+p phases can be avoided by first making a continuous transition to the d-wave phase and another continuous transition to the chiral d+p phase by suitably tuning $U_0$ and $\nu$.

When examining the gap solution as a function of $\bm k$, we find cusps when approaching the Fermi surface in all phases. In the chiral phase d+p phase, the p-wave contribution exhibits peaks at the former location of the d-wave nodes. This fast variation of the solution along the Fermi surface becomes more and more pronounced as the range of the interaction and its strength increase. The cusps in the gap solution are likely to alter the transport and thermodynamic properties of the superconductor. Another publication will be devoted to their study.   

\subsection{Tunable mode stability in non-equilibrium \\ Higgs spectroscopy}

After studying exotic emergent phases due to long-range interactions in the last section,
we now evaluate the impact of these interactions on the dynamical properties of the superconductor.
In the following, we excite Higgs oscillations of the condensate through a sudden quench and analyze the arising non-equilibrium dynamics. 

Detecting the faint signal of the Higgs mode in superconductors is a challenging experimental task, as the mode does not couple linearly to electromagnetic fields. Nevertheless, Higgs spectroscopy holds large promise. It has been recently shown that the spectrum of the Higgs mode can reveal the symmetry properties of unconventional superconductors \cite{schwarz2020classification},  allowing for a deeper understanding of how exotic superconductors acquire their properties.

\begin{figure}
    \centering
    \includegraphics[width=.47\textwidth]{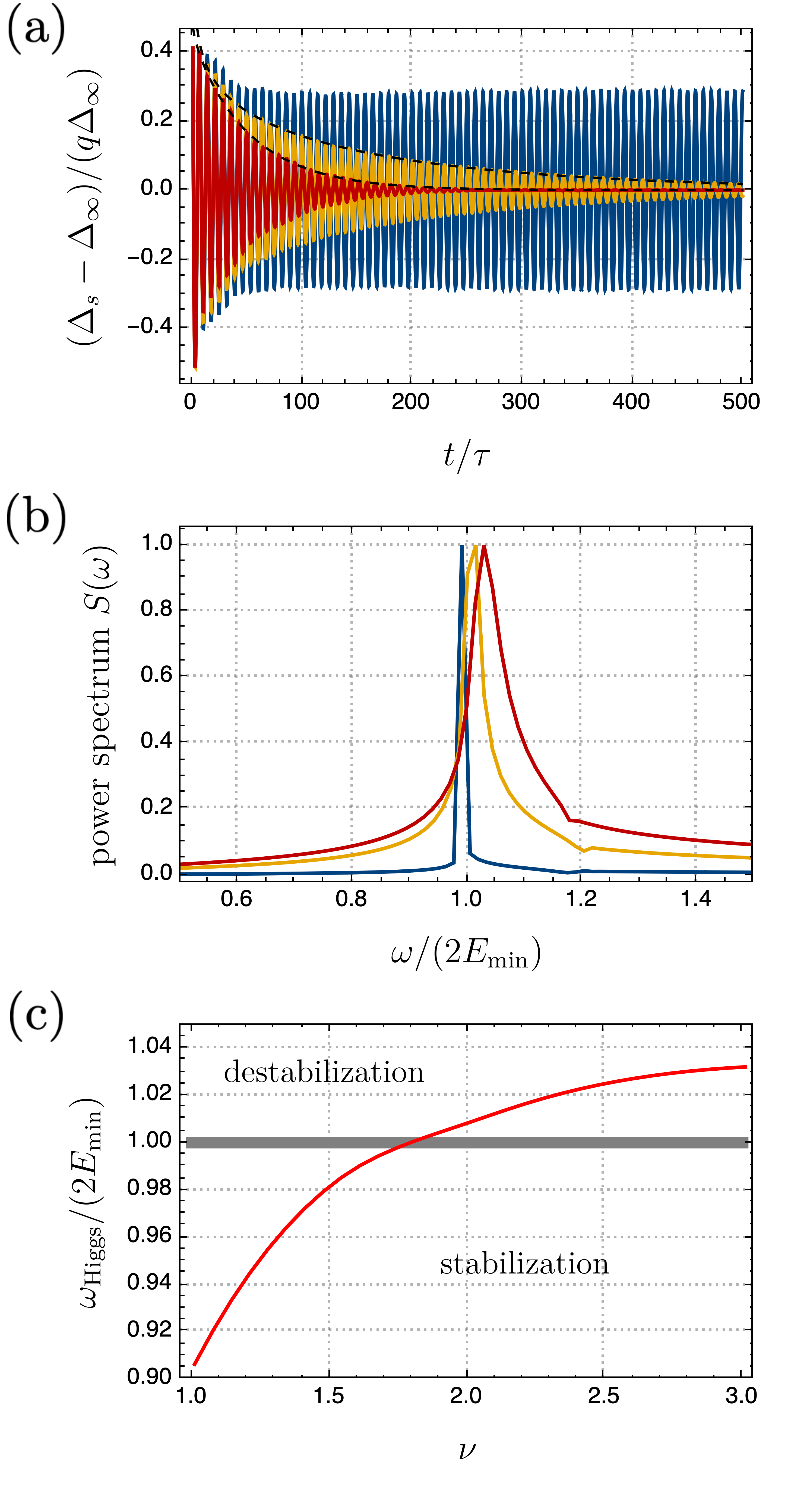}
    \caption{\textcolor{myred}{(a) Higgs oscillation of a 2D superconductor in the s-wave phase for $C_0=1.5$, $U_0=1$ for the exponents $\nu = 1.61$ (blue), $\nu = 2.01$ (yellow), and $\nu=3.01$ after suddenly reducing $C_0$ and $U_0$ by a quench factor $(1-q)$ with $q=10^{-2}$. The blue curve shows Higgs stabilization due to long-range interactions, while the yellow and red curves display exponential decay of the oscillation amplitude. (b) Power spectrum of the Higgs oscillation with parameters as in (a). (c) Higgs mode frequency in units of the minimum dispersion relation after the quench $E_\text{min}$ as a function of $\nu$ (red). Above the gray line, the Higgs oscillation decays exponentially, whereas below, it is stabilized.}}
    \label{fig:higgs}
\end{figure}

We now perform a sudden quench in the s-wave phase with $C_0=1.5$ and $U_0=1$, changing $C_0\to (1-q) C_0$, $U_0\to (1-q)U_0$ with a quench parameter $q=10^{-2}$. We then determine the time evolution of the condensate and evaluate the s-wave order parameter $\Delta_s$ defined as
\[
\Delta_s = V_\Lambda \bigg\vert \int_{E^\ast} \Delta_{\uparrow \downarrow}(\bm k) \,\mathrm d \bm k\bigg\vert.
\]
The resulting Higgs oscillations of $\Delta_s$ are displayed in Fig.~\ref{fig:higgs}~(a) for $\nu=1.61$ (blue curve), $\nu=2.01$ (yellow), and $\nu = 3.01$ (red). Here we have centered the curves around the mean value of the oscillation $\Delta_\infty$. We find that, for $\nu = 3.01$ and $\nu =2.01$, the long-range interactions lead to a fast decay of the amplitude $a(t)$ of the Higgs oscillation where both algebraic and exponential contributions are present, with
\[
a(t)\sim t^{-\alpha} e^{-\beta t}.
\] 
The black dashed curves in Fig.~\ref{fig:higgs}~(a) display the corresponding fits to the oscillation maxima. In contrast to the standard $t^{-1/2}$ decay \cite{PhysRevLett.96.097005}, which we recover for $U_0=0$, the long-range interaction can destabilize the Higgs mode. As the range of the interaction is increased, the decay coefficient $\beta$ decreases monotonically. As soon as $\nu$ becomes smaller than a certain threshold value $\nu_\mathrm{crit}$, the long-range interactions stabilize the Higgs mode as displayed in the blue curve for $\nu=1.61$.

In Fig.~\ref{fig:higgs}~(b), we display the power spectrum of the oscillation. Here the frequency is shown in units of $2 E_\text{min}$, where $E_\text{min}$ is the minimum quasi-particle dispersion relation
\[
 E_\text{min}=\min_{\bm k} E(\bm k)
\]
and where $2 E_\text{min}$ coincides with the Higgs oscillation frequency for vanishing interaction ($U_0=0$). The broadening of the spectral peaks reflects the increasing damping of the Higgs mode as $\nu$ increases.

Recently, it has been found that Higgs oscillations can be stabilized by infinite-ranged interactions ($\nu = 0$) \cite{gao2021higgs}, leaving the question unanswered of what effect realistic interactions with $\nu>0$ have on the Higgs mode's decay. Using our toolset, we find that with long-range interactions, the decay of the Higgs mode can be tuned. For sufficiently large $\nu$, the decay of the Higgs mode is accelerated compared to the well-known $t^{-1/2}$ decay and includes both algebraic and exponential contributions. Only below a certain value of $\nu$, the Higgs mode becomes stable. The origin of the tunability of the Higgs mode's stability is explained in Fig.~\ref{fig:higgs} (c). We show the  Higgs frequency as a function of the interaction exponent $\nu$. Destabilization of the Higgs mode occurs, as soon as the Higgs mode's frequency exceeds the minimum frequency of the quasi-particle spectrum (gray line). The mode is pushed into the quasi-particle spectrum in contrast to the case of no long-range interaction ($U_0=0$), where the Higgs mode merely lies at its boundary. Being now placed inside this spectrum, the coupling to the quasi-particles is enhanced, which creates additional exponential damping. If $\nu$ is smaller than the threshold value, Higgs mode and quasi-particle spectrum are pushed apart, and a gap opens between them. This leads them to become decoupled, and hence the Higgs mode remains stable.

\color{black}
 
\section{Conclusions and Perspectives}
\label{sec:conclusions}

Long-range interacting systems, both on a lattice and in the continuum, are highly relevant as they transcend our understanding of short-range physics. Among others, it is well-known that these systems can exhibit non-local correlations that can alter critical exponents continuously, driving the system to completely novel phases with many open questions \cite{fey2019quantum,PhysRevB.96.104432,PhysRevLett.121.090603,PhysRevB.93.125128,PhysRevA.93.053620,PhysRevB.102.174424,PhysRevB.97.155116,zaletel2015time,PhysRevLett.121.090603,Saito2020,kuwahara2020area}. They are actively being explored in experiment \cite{Britton2012,Schauss2012,PhysRevLett.108.210401,Yan2013,Richerme2014,Douglas2015,Landig2016} with potential applications in quantum computing and quantum simulation \cite{Martinez2016,RevModPhys.93.025001,Scholl2021}.  

The problem of establishing the connection between the long-range interacting lattice problem and the associated continuum field theory has so far only been approached for specific systems, and the validity of this connection has often remained questionable. Among others, arbitrary ultraviolet cutoffs need to be introduced in order to make the field theory well--defined, hence introducing free parameters in theory. Furthermore, the continuum limit tends to break down in the case where the interaction exponent matches the system dimension, e.g., in long-range interacting quantum magnetic systems, where new types of quantum phases and phase transitions have been conjectured \cite{maghrebi2017continuous}.

This work solves the problem stated above and establishes the previously elusive connection for a very broad set of physical systems. For any number of space dimensions, for any lattice with any number of atoms per unit cell, for any power-law interaction, and both for linear and nonlinear systems,  we provide an exact representation of long-range interacting lattice problems in terms of their associated continuum theories and vice versa. This representation can then either be used as an analytical tool or as a numerical method aiming at advancing our understanding of the critical behavior of both quantum and classical systems with long-range interactions. Provided that the function $g$ describing the properties of the lattice, e.g., spin or particle displacement, varies sufficiently slowly, the lattice problem can be separated into a continuum contribution and a lattice contribution. The lattice contribution can be written in terms of a differential operator that is based on the Epstein zeta function, the generalization of the Riemann zeta function to multi-dimensional oscillatory lattice sums.
Along with this article, we provide an implementation of Epstein zeta for arbitrary lattices in the supplemental material as well as on GitHub~\cite{Buchheit_Continuum_representation_2022}. Using finite order approximations to this differential operator, we are able to compute singular lattice sums with excellent precision and at the numerical cost of an integral approximation. 

We benchmark our method, computing energies or forces in three physical examples:
studying Skyrmions in a 2D classical Heisenberg spin lattice with dipolar interactions,
kinks in an ion chain with the nonlinear Coulomb interaction,
as well as spin waves in a three-dimensional pyrochlore lattice with dipole interactions. In all three cases, our representation yields an excellent agreement with exact summation in contrast to the standard integral approximation. We show that the lattice contributions are needed in order to obtain reliable results, whereas, in the case of the pyrochlore lattice, the continuum approximation even fails in reproducing the correct qualitative behavior.

\textcolor{myred}{
As the key application of our method, we solve the important problem of understanding unconventional superconductors with long-range interactions. Using our representation in Fourier space, we derive a generalized BCS equation for the superconducting gap matrix that is valid for all power-law interactions and both singlet and triplet pairing. The Epstein zeta function, which encodes the information about the microscopic structure of the material, is of critical importance here and enters as an integral kernel in the gap equation. We demonstrate that including long-range interactions leads to a rich phase diagram with unconventional s, d, d+s, and d+p phases separated by quantum-critical continuous transitions. In the d+s and d+p phases, the superconductor chooses an exotic pairing structure to avoid energetically costly nodes in the quasi-particle dispersion relation. Here, the d+p phase is topologically nontrivial. As the strength and range of the long-range interaction increase, the superconducting gap forms sharp features along the Fermi surface. The implication of these features on the physical properties of the superconductor, such as transport properties, remains an open question that will be answered in future publications. We subsequently analyze the impact of long-range interactions on the non-equilibrium behavior of superconductors in the s-wave phase by spectroscopy of Higgs oscillations after a sudden quench. In contrast to previous works, we show that the interactions can tune the decay behavior of the Higgs mode's amplitude. For interactions that fall off sufficiently fast, the interactions lead to an exponential decay of the oscillations, whereas sufficiently long-ranged interactions stabilize the mode. This behavior arises due to the Higgs mode either being incorporated into the quasi-particle spectrum, increasing their coupling, or being separated from it, removing their coupling. 
}

\textcolor{myred}{
Applying the continuum representation and the SEM to unconventional superconductivity opens up numerous possibilities for further investigations. The properties of the arising exotic phases and the impact of long-range interactions on topological properties and edge states are particularly interesting. Using our representation in real space will allow investigating superconductors with impurities, vortices, or boundaries. A noteworthy area of research is the study of Majorana modes in vortices and their interplay with long-range interactions.}

\textcolor{myred}{
We have shown the first example of Higgs spectroscopy with realistic long-range interactions in the s-wave phase. Further investigations will examine the non-equilibrium behavior of the condensate's oscillations in the exotic phases. Here, the interplay between the different symmetries and their oscillation modes is of interest.
}

\textcolor{myred}{
    Looking further, we can use our representations on any quantum system on a lattice within the mean field approximation. Direct applications to magnetic systems, where long-range interactions can be created and tuned in THz nanoplasmonic cavities, come to mind. We consider it worthwhile to investigate the quantum critical behavior of long-range Ising chains in a transverse field, especially when the interaction exponent equals the system dimension where new phases of matter are being expected but where standard approaches reach their limits \cite{maghrebi2017continuous}. 
}

\textcolor{myred}{On the methodical side, different extensions of our representation are possible. For finite systems, geometry-dependent terms arise, which can be described within our method as well \cite{buchheit2022singular}. These terms can be of relevance, e.g., in mesoscopic systems or in quantum-Hall type topological materials that can exhibit soliton-like edge states \cite{mukherjee2021observation}. Finally, we aim to combine the continuum representation in Fourier space with diagrammatic methods, such as the T-matrix approach \cite{stefanucci2013nonequilibrium}, to investigate quantum systems beyond the mean field approximation.
}

\section*{Acknowledgments}

We thank Frank K. Wilhelm, David E. Bruschi, and G\"otz S.\ Uhrig for fruitful discussions. \textcolor{myred}{The authors gratefully acknowledge the scientific support and HPC resources provided by the Erlangen National High Performance Computing Center (NHR@FAU) of the Friedrich-Alexander-Universit{\"a}t Erlangen-N{\"u}rnberg (FAU) under the NHR project n101af. In particular, we thank Thomas Gruber for providing excellent code review. NHR funding is provided by federal and Bavarian state authorities. NHR@FAU hardware is partially funded by the German Research Foundation (DFG) – 440719683.
TK acknowledges funding received from the European Union’s Horizon 2020 research and innovation programme under the Marie Skłodowska-Curie grant agreement No 899987.
\begin{center}
\includegraphics[width=0.35\linewidth]{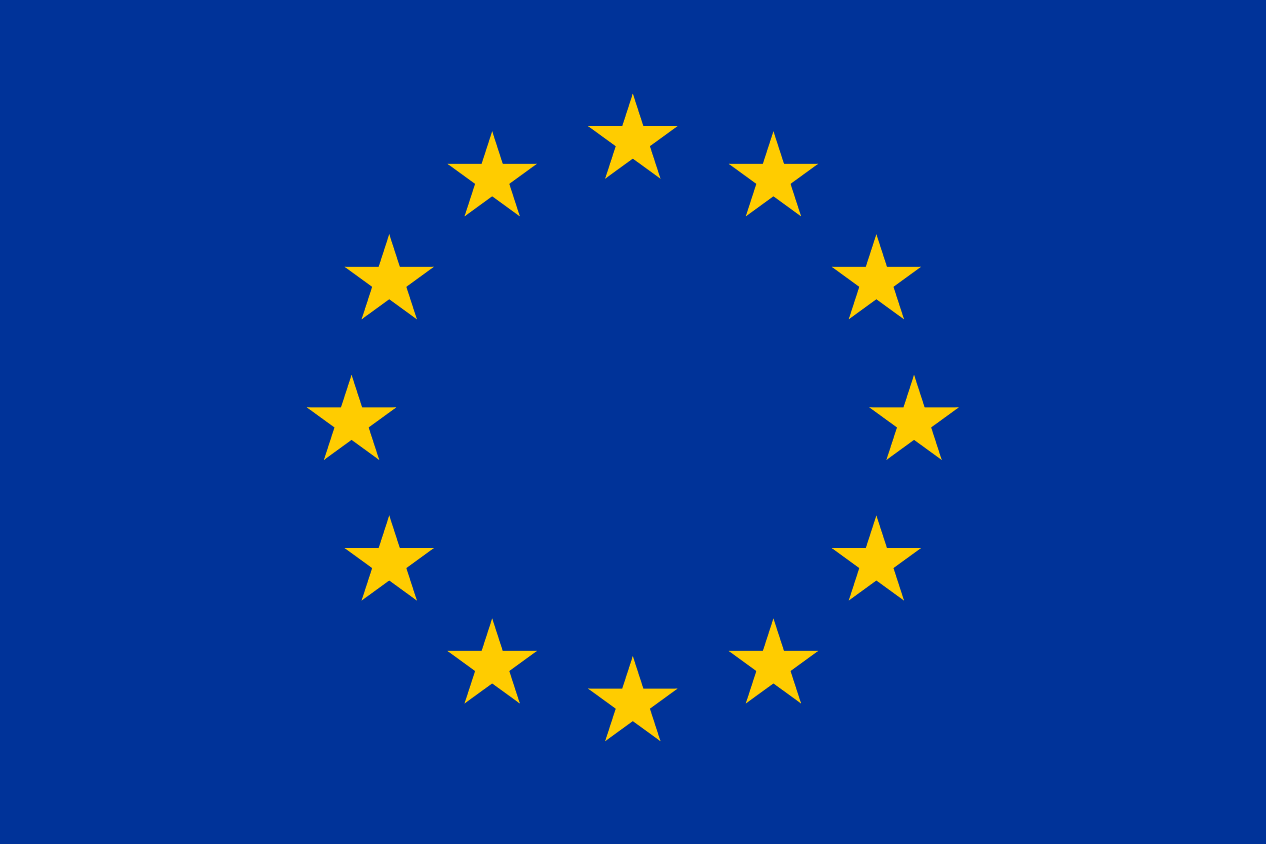}
\end{center}
}

\appendix
\section{Hadamard integral}
\label{sec:appendix_hadamard}
The Hadamard finite-part integral is the natural extension of the standard integral to functions that exhibit non-integrable power-law singularities,
see the original publication of Hadamard~\cite{hadamard1952lectures} or Ref.~\cite{gelfand1964generalizedI}.
Here, we follow the notation of Ref.~\cite{thesis_buchheit_2021}.
For
\[
f_{\bm x}(\bm y) = \frac{g(\bm y)}{\vert \bm y-\bm x\vert^\nu},
\]
with $g$ sufficiently differentiable,
the Hadamard integral of $f_{\bm x}$ over a domain $\Omega$ is defined by subtracting the Taylor series of $g$ up to order 
\[
    k_\text{max}=\lfloor \mathrm{Re}(\nu)-d \rfloor,
\] 
with $\lfloor x \rfloor$ the largest integer smaller than or equal to $x$. The Hadamard integral reads
  \[
  \ddashint \limits_\Omega  f_{\bm x}(\bm y)  \,\mathrm d \bm y =\lim_{\varepsilon\to 0}\Bigg(\,\int \limits_{\Omega\setminus B_{\varepsilon}(\bm x)}  f_{\bm x}(\bm y)  \,\mathrm d \bm y-\big(\mathcal H_{\nu,\varepsilon}g\big)(\bm x)\Bigg),
  \]
with the differential operator
\[
  \mathcal H_{\nu,\varepsilon}=\sum_{k=0}^{k_\text{max}} \frac{1}{k!} \int\limits_{\mathds R^d\setminus B_\varepsilon} \frac{(\bm y \cdot \bm \nabla)^k}{\vert \bm y \vert^\nu}\,\mathrm d \bm y,\quad \nu \in \mathds C\setminus (\mathds N+d).
\]
For $\text{Re}(\nu)<d$, the Hadamard integral coincides with the standard integral, otherwise, it forms its meromorphic continuation in $\nu$. For the special cases $\nu \in (\mathds N+d)$, we can define the Hadamard integral uniquely up to derivatives of the function $g$ of order $\nu-d$. We can choose
\begin{align*}
  \mathcal H_{\nu,\varepsilon}&=\sum_{k=0}^{k_\text{max}-1} \frac{1}{k!} \int\limits_{\mathds R^d\setminus B_\varepsilon} \frac{(\bm y \cdot \bm \nabla)^k}{\vert \bm y \vert^\nu}\,\mathrm d \bm y \notag \\ &+  \frac{1}{k_\text{max}!} \int\limits_{B_1\setminus B_\varepsilon} \frac{(\bm y \cdot \bm \nabla)^{k_\text{max}}}{\vert \bm y \vert^\nu}\,\mathrm d \bm y.
\end{align*}
Other choices for the Hadamard integral for these special cases can be obtained by replacing the ball $B_1$ with a sufficiently regular neighborhood of $\bm y=\bm 0$.

\section{Numerical integration}
\label{sec:num-integration}
For the computation of the Hadamard integrals appearing in this work, we first use spherical coordinates
to split the integration over $\mathds R^d$ into a nonsingular integral over the unit sphere $S^{d-1}$
and a singular radial integral. The goal is now to approximate the integral by a finite sum of suitably weighted point evaluations of the integrand so that the error falls off exponentially with the number of points. This particular choice of evaluation points and weights is called the numerical integration (or quadrature) rule. In the following, the desired convergence is achieved by combining trapezoidal and Gauss quadrature rules~\cite{stoer2002introduction}.  The integral over the unit sphere is computed by the trapezoidal rule
for $d = 2$ and a tensor product of trapezoidal and Gauss rules for $d = 3$.
The radial integral is computed by a specialized Gauss quadrature.
For the general case of
\[
\ddashint_0^\infty r^{-\nu + d - 1} g(r) \, \text d r 
\]
with a quickly decaying function $g$, we first restrict the integration domain to the finite interval $[0, R]$, assuming that the integrand falls off fast enough such that the integral over $[R,\infty)$ can be neglected.
By the definition of the Hadamard integral in Sec.~\ref{sec:appendix_hadamard}
we can express the above integral as an ordinary integral, provided we subtract the
Taylor expansion $p$ of $g$ with sufficiently high order,
\[
I = \int_0^R r^{-\nu + d - 1} \big( g(r) - p(r) \big) \, \text d r
\]
Now, $g(r) - p(r) = r^{k_\text{max} + 1} h(r)$ with $h(r)$ bounded as $r \to 0$, so $I$ is an integral over $h$
with integrable weight.
After a change of variables,
\[
I = c \int_0^1 r^{-\nu + d + k_\text{max}} h( R r) \, \text d r,
\]
with $c=R^{-\nu + d +k_\text{max} +1}$.
For the computation of this integral, we employ the Gauss-Jacobi rules,
exponentially convergent quadrature rules for integrals of the form
\[
\int_0^1 r^\alpha f(r) \, \text d r
\]
with $\alpha > -1$.
The nodes $r_j$ and weights $w_j$, $j=1,\dots,m$, of this scheme for fixed $\alpha$ can be efficiently precomputed and stored. An implementation of the associated algorithm can be found in our code, which is provided in the supplemental material and on our GitHub repository~\cite{Buchheit_Continuum_representation_2022}. Choosing $\alpha = - \nu + d + k_\text{max}$, the integral is computed via
\begin{align*}
I \approx c \sum_{j = 1}^m w_j h(R r_j)
&= c \sum_{j = 1}^m \frac{w_j}{r_j^{k_\text{max}+1}} \big( g(R r_j) - p(R r_j) \big),
\end{align*}
thus obtaining a quadrature rule for the Hadamard integral
that involves point evaluations of $g$ and its derivatives.

\section{Pyrochlore lattice structure}
\label{sec:app_pyrochlore}

The lattice vectors of the pyrochlore lattice structure are given by
\[
\bm a_1=-\left(\begin{matrix}1\\ \sqrt{3}\\ 0  \end{matrix}\right),~ \bm a_2=\left(\begin{matrix}1\\ -\sqrt{3}\\ 0  \end{matrix}\right),~ \bm a_3=\left(\begin{matrix}1\\ -2/\sqrt{3}\\ 2\sqrt{2/3}  \end{matrix}\right),
\]
and the positions of the $n=4$ atoms in the elementary lattice cell read $\bm d_1=\bm 0$,
\[
    \bm d_2 = \left(\begin{matrix}-1\\ 0\\ 0  \end{matrix}\right),\,\bm d_3 = -\left(\begin{matrix}1/2\\ \sqrt{3}/2\\ 0  \end{matrix}\right),\,\bm d_4 =- \left(\begin{matrix}1/2\\ 1/(2\sqrt{3})\\ \sqrt{2/3}  \end{matrix}\right),
\]
see Ref.~\cite{siddharthan2001spin}. 

\section{Anomalous quantum spin wave dispersion}
\label{sec:anomalous_dispersion}
We now determine the quantum dispersion relation for linear spin waves in an $n$-atomic lattice with ferromagnetic long-range interactions. The system Hamiltonian  reads
\begin{align*}
    H&=-\frac{J}{2} \sideset{}{'}\sum_{i,j=1}^n\, \sum_{\bm x,\bm y\in \Lambda} \frac{\bm S_{\bm x,i}\cdot \bm S_{\bm y,j}}{\vert  (\bm x+\bm d_i)-(\bm y+\bm d_j) \vert^\nu} 
\end{align*}
with $\bm S_{\bm x,i}$ the spin operator for the site  $\bm x+\bm d_i$.
The scalar product of the spin operators can be conveniently written in the representation
\[
\bm S_{\bm x,i}\cdot \bm S_{\bm y,j}=S_{\bm x,i}^z S_{\bm y,j}^z+(S_{\bm x,i}^+ S_{\bm y,j}^-+S_{\bm x,i}^- S_{\bm y,j}^+)/2,
\]
with $S_{\bm x,i}^+=S_{\bm x,i}^x+i S_{\bm x,i}^y$ and $S_{\bm x,i}^-=S_{\bm x,i}^x-i S_{\bm x,i}^y$.
Under the standard Holstein--Primakoff transformation, the spin operators are cast in terms of the bosonic creation and annihilation operators $a_{\bm x,i}^\dagger$ and $a_{\bm x,i}$, 
\begin{align*}
    S_{\bm x,i}^z&=-S+a_{\bm x,i}^\dagger a_{\bm x,i}, \quad S_{\bm x,i}^+=a_{\bm x,i}^\dagger \sqrt{2S-a_{\bm x,i}^\dagger a_{\bm x,i}},\notag \\ S_{\bm x,i}^- &=(S_{\bm x,i}^+)^\dagger.
\end{align*}
In the large $S$ limit, restricting the Hilbert space to states where $\langle a_{\bm x}^\dagger a_{\bm x}\rangle /S\ll 1$, we can replace the Hamiltonian by
\begin{align*}
H= J S\sum_{i,j=1}^n\, \sideset{}{'}\sum_{\bm x,\bm y\in \Lambda}  \frac{a_{\bm x,i}^\dagger a_{\bm x,i}+a_{\bm x,i}^\dagger a_{\bm x+\bm y,j}}{\vert \bm y-(\bm d_i-\bm d_j) \vert^\nu},
\end{align*}
where we have discarded the constant ground state energy.
Subsequently, we write the annihilation operators for each sublattice in Fourier space \[
a_{\bm x, i}= \sqrt{V_{\Lambda}}\int \limits_{E^\ast} e^{2\pi i \bm k\cdot (\bm x+\bm d_i)} a_{\bm k,i}\,\mathrm d\bm k
\]
with $E^\ast={(A_\Lambda^{-1})^T[-1/2,1/2]^d}$ the first Brillouin zone. Using that 
\[
    V_\Lambda \sum_{\bm x\in \Lambda } e^{-2\pi i\bm k\cdot \bm x} =\delta_{\bm k},\quad \bm k\in E^\ast,
\]
with $\delta_{\bm k}$ the Dirac delta distribution, we then find that 
\[
    H= \int \limits_{E^\ast} \sum_{i,j=1}^n \big(B(\bm k) \big)_{i,j} a_{\bm k,i}^\dagger a_{\bm k,j}\,\mathrm d\bm k
\]
with the Hermitian matrix $B(\bm k)\in \mathds R^{n\times n}$,
\begin{align*}
    \big( B(\bm k)\big)_{i,j} &= J S\Bigg( \sum_{m=1}^n \delta_{i,j} Z_{\Lambda}\left\vert \begin{matrix}
      \bm d_i-\bm d_m\\\bm 0
    \end{matrix} \right\vert \notag \\ 
    &- e^{2\pi i (\bm 
    d_i-\bm d_j)\cdot \bm k}Z_{\Lambda}\left\vert \begin{matrix}
      \bm d_i-\bm d_j\\ \bm k
    \end{matrix} \right\vert\Bigg), 
\end{align*}
with $\delta_{i,j}$ the Kronecker delta.
In an $n$-atomic lattice, the spectrum of the Hamiltonian $H$ then exhibits $n$ bands, which follow from the eigenvalues of $B$,  
\[
\hbar \omega_i(\bm k) = \text{Eig}\big(B(\bm k)\big)_i,\quad i=1,\dots, n,
\]
with $\text{Eig}$ the vector of eigenvalues and where $\hbar \omega_i$ are the band energies. In the case of a mono-atomic lattice, this reduces to the simple relation 
\[
\hbar \omega(\bm k) = JS \Bigg( Z_{\Lambda}\left\vert \begin{matrix}
      \bm 0\\\bm 0
    \end{matrix}\right \vert(\nu) - Z_{\Lambda}\left\vert \begin{matrix}
      \bm 0\\\bm k
    \end{matrix} \right \vert (\nu)\Bigg).
\]
In the long-wavelength limit, we obtain
\[
    \hbar\omega(\bm k) \approx JS \Bigg( \frac{\hat s(\bm k)}{V_\Lambda} - \frac{1}{2}(\bm k \cdot \bm \nabla_{\bm y})^2 Z_{\Lambda}^\text{reg}\left\vert \begin{matrix}
      \bm 0\\\bm y
    \end{matrix} \right \vert (\nu)\bigg\vert_{\bm y=0}\Bigg),
\]
with corrections of order $\mathcal O(\bm k^4)$.
Here the typical $\mathcal O(\bm k^2)$ scaling is observed in case that $\text{Re}(\nu)>d+2$. On the other hand, for $\text{Re}(\nu)< d+2$, the Fourier transform of the interaction, cf. Eq.~\eqref{eq:fourier-interaction}, dominates and we have
\[
\omega(\bm k) \sim \vert \bm k\vert^{\nu-d},
\]
leading to an anomalous dispersive behavior of the spin-lattice due to the long-range interaction. Our result can be applied to lattices in any dimension and for any interaction exponent $\nu$. In particular, it generalizes results that have previously obtained for $d=1$ in Ref.~\cite{yusuf2004spin} where anomalous behavior was predicted for an antiferromagnetic spin chain with long-range interactions $1<\text{Re}(\nu)<3$ \color{myred}
as well as similar results in the long-range transverse Ising chain \cite{PhysRevLett.111.207202} \color{black}. Furthermore, it captures the linear scaling of the dispersion relation observed in a $d=2$ spin lattice with dipolar interactions in Ref.~\cite{peter2012anomalous}.

\color{myred}
\section{Bogoliubov transformation}
\label{sec:bogoliubov_transform}
In this section, we provide details on the diagonalization of the Hamiltonian in Sec.~\ref{sec:BCS_ground_state}.
Using the singular value decomposition of $\Delta$, namely
\[
\Delta = U \Sigma V^\dagger,
\]
with $U,V\in \mathds C^{2\times 2}$ unitary and $\Sigma=\text{diag}(\sigma_1,\sigma_2)$  the diagonal matrix that contains the non-negative singular values $\sigma_i$ of $\Delta$, we find the energy bands 
\[
E_i(\bm k)=\sqrt{\xi^2(\bm k)+\sigma_i^2(\bm k)}.
\]
From the gap matrix symmetry $\Delta(\bm k)=-\Delta^T(-\bm k)$, it follows that  $\sigma_i$ and hence $E_i$ are centro-symmetric
\[
\sigma_i(-\bm k) = \sigma_i(\bm k),\quad  E_i(\bm k)=E_i(-\bm k).
\]
After setting 
\begin{align*}
    u_{i}=\sqrt{\frac{1}{2}\left(1+\frac{\xi}{E_i}\right)},\quad v_i=\sqrt{\frac{1}{2}\left(1-\frac{\xi}{E_i}\right)},
\end{align*}
and $\bm u=(u_1,u_2)^T$, $\bm v=(v_1,v_2)^T$,
we find that the matrix
\begin{align*}
B = B_0 B_1 B_2&=  \left( \begin{matrix} U & 0 \\ 0 & V
\end{matrix} \right)\left( \begin{matrix} \text{diag}(\bm u) & -\text{diag}(\bm v) \\ -\text{diag}(\bm v) & -\text{diag}(\bm u)
\end{matrix} \right)  \\ &\times \left(\begin{matrix}
\mathds 1_2 & 0 \\ 0 & U^\dagger(\bm k)  V^*(-\bm k)
\end{matrix}\right)
\end{align*}
diagonalizes $\mathcal H$. Here the matrix $B_0$ renders $\mathcal H$ block-diagonal and real,
\[
B_0^\dagger \mathcal H B_0 = \left( \begin{matrix} \xi \mathds 1_2 & -\Sigma \\ -\Sigma & -\xi \mathds 1_2
\end{matrix}\right),
\]
the second matrix diagonalizes $\mathcal H$,
while the third matrix ensures that $\bm \gamma$ only includes two fermionic operators $\gamma_1$ and $\gamma_2$ where
\[
\bm \gamma (\bm k) = \Big(\gamma_{1}(\bm k), \gamma_{2}(\bm k), \gamma_1^\dagger(-\bm k), \gamma_2^\dagger(-\bm k) \Big)^T=B^\dagger(\bm k) \bm \Psi(\bm k),
\]
while leaving the diagonal form of $\mathcal H$ unchanged.
The full transformation reads 
\[
B^\dagger\mathcal H B =  \text{diag}\big(E_1,E_2,-E_1,-E_2\big).
\]
The Hamiltonian then reads 
\begin{align*}
H= \frac{1}{2}\sum_{i=1}^2\int \limits_{E^*}  &E_i(\bm k)\gamma_{i}^\dagger(\bm k)\gamma_i(\bm k)\\- &E_i(\bm k)\gamma_{i}(-\bm k) ) \gamma_i^\dagger(-\bm k) \,\mathrm d \bm k.
\end{align*}
After discarding a constant term, we hence find
\[
H=  \sum_{i=1}^2 \int \limits_{E^*} E_i(\bm k)        \gamma_i^\dagger (\bm k)\gamma_i (\bm k) \,\mathrm d \bm k,
\]
and the BCS ground state is the vacuum in the new operator basis
\[
\vert \psi_\text{BCS}\rangle = \vert \text{vac}\rangle.
\]
After inserting the time evolution in the basis of the original fermionic operators
\[
\bm \Psi (\bm k) =  U(\bm k) B (\bm k) \bm \gamma(\bm k),
\]
into the definition of the density matrix
\[
\rho_{i,j}(\bm k) = \frac{1}{2} \ddashint_{E^*} \langle \psi_\text{BCS}\vert \Psi^\dagger_j(\bm k') \Psi_i(\bm k) \vert \psi_\text{BCS}\rangle \,\mathrm d \bm k'
\]
we find that
\[
\rho(\bm k,t) =  B(\bm k) \left(\begin{matrix}   0 & 0  \\ 0 & \mathds 1_2   \end{matrix}\right) B^\dagger(\bm k) ,
\]
where the above zeros denote the $2\times2$ zero matrix. Then
\begin{align*}
\rho= \frac{1}{2}\left(
\begin{matrix} 
U\text{diag}(\bm v)^2 U^\dagger & U \text{diag}(\bm u)\text{diag}(\bm v) V^\dagger \\
V^\dagger \text{diag}(\bm u)\text{diag}(\bm v) U & V\text{diag}(\bm u)^2 V^\dagger
\end{matrix}
\right).
\end{align*}
After noting that 
\[
u_i^2 = \frac{1}{2}+\frac{\xi}{2E_i},\quad v_i^2 = \frac{1}{2}-\frac{\xi}{2E_i},\quad u_i v_i=\frac{\sigma_i}{2 E_i},
\]
the density matrix $\rho$ follows as
\[
\frac{1}{4}\left(
\begin{matrix} 
\mathds 1_2-U \text{diag}\Big(\frac{\xi}{E_1},\frac{\xi}{E_2}\Big) U^\dagger & U \text{diag}\Big(\frac{\sigma_1}{E_1},\frac{\sigma_2}{E_2}\Big) V^\dagger \\
V \text{diag}\Big(\frac{\sigma_1}{E_1},\frac{\sigma_2}{E_2}\Big) U^\dagger & \mathds 1_2+V\text{diag}\Big(\frac{\xi}{E_1},\frac{\xi}{E_2}\Big)V^\dagger
\end{matrix}\right),
\]
which can be brought in the compact form
\[
\rho= \frac{1}{4}\left(
\begin{matrix} 
\mathds 1_2 -\xi E[\Delta^\dagger]^{-1} & \Delta E[\Delta]^{-1}\\
\Delta^\dagger E[\Delta^\dagger]^{-1} & \mathds 1_2 + \xi E[\Delta]^{-1}
\end{matrix}\right).
\]
Collecting all constant energy contributions allows us to compute the ground state energy as 
\begin{align*}
E_\text{GS}
= & \frac{N}{2}  \sum_{\sigma\sigma'} \int \limits_{E^\ast} V_\Lambda \int \limits_{E^\ast } \Big(C_{\sigma\sigma'}+U_{\sigma\sigma'} Z_{\Lambda,\nu} \left \vert \begin{matrix} 0 \\ \bm q \end{matrix} \right \vert\Big)
\\  &\times  \alpha_{\sigma\sigma'}(\bm k-\bm q) \alpha^*_{\sigma\sigma'}(\bm k)\,\mathrm d \bm q\,\mathrm d \bm k \notag \\ &+\frac{N}{2} \int \limits_{E^\ast} \text{Tr}\big(\xi \mathds 1_2-E[\Delta]\big)(\bm k)\,\mathrm d \bm k,
\end{align*}
with $N$ an extensive constant proportional to the number of sites.

\section{Numerical simulation of unconventional superconductors}
\label{sec:numerics}

The numerical simulation of unconventional superconductors with long-range interactions poses several
numerical challenges whose solutions we shortly address here.
First, the computation of the gap $\Delta$ via~\eqref{eq:generalized_gap_equation}
requires the computation of the Epstein zeta function $Z_{\Lambda,\nu}$ on a grid in momentum space.
To that end, we employ the representation of $Z_{\Lambda,\nu}$ in~\cite{crandall2012unified} as an exponentially
fast converging series.
This requires the computation of the incomplete gamma function for which use the recently
developed algorithm in~\cite{abergel2020algorithm}.

Numerical integration of the singular continuum contribution is performed
by first dividing the Brillouin zone into triangles.
Using the nonlinear Duffy transform~\cite{duffy1982quadrature}
we change the domain of integration to the unit square $[0,1]^2$.
The typical power law singularity \[\sqrt{k_1^2 + k_2^2}^\nu\]
factorizes under the Duffy transform $\bm k = (u, u v)^T$
into a singular and an analytical part,
\[
\sqrt{k_1^2 + k_2^2}^\nu = u^\alpha \sqrt{1 + v^2}^\alpha.
\]
For the integration of the singular part, we use the techniques
described in Appendix~\ref{sec:num-integration}.
The regular part is integrated by an exponentially convergent
Gau{\ss}--Legendre rule.

In the case of the stationary gap equation, we observe that for $C_0=0$ a fixed-point iteration
starting with a random initial guess reliably converges to the ground state up to
machine precision. Bistability occurs for finite $C_0$ close to the first-order transition. Here the ground state is found by  comparing the energies of an s-wave-biased solution and a d-wave-biased solution, choosing the one with the lowest energy.

The result is validated by restarting the fixed-point iteration with an initial
guess biased towards s-, p-, or d-symmetry and comparing the computed ground state energies.

At every time step, the quasi-particle energy 
needs to be computed at every grid point.
For an efficient and stable computation via the singular value
decomposition of $\Delta$ we use the specialized
routine described in~\cite{novakovic2020batched}.

Finally, the numerical solution of the time evolution
equation requires that the propagator $U_t$ remains unitary
at all time steps. Usual Runge--Kutta methods can not
guarantee this.
Therefore, we use the Lie group methods of
Crouch and Grossman~\cite{crouch1993numerical}
that, by construction, guarantee that the numerical solution
to the time evolution equation will always be unitary.
The second-order scheme reads
\[
U_{t+\delta t} = \exp \Big(-i\delta t \mathcal H\Big[\exp\Big(-i\frac{\delta t}{2} \mathcal H[U_t]\Big) U_t\Big]\Big) U_t,
\]
where desired convergence is achieved by first making a prediction for the time evolution operator at a half step $\delta t/2$ and then using this predictor to generate the full time step $\delta t$.
Higher order methods up to order five are presented in~\cite{jackiewicz2000construction,marthinsen2001note}.

\color{black}

\end{document}